\documentclass[aps,preprint,amssymb,12pt,floatfix]{revtex4}
\usepackage{subfigure}
\usepackage[pdftex]{graphicx}
\usepackage{rotating}
\usepackage{epstopdf}
\usepackage{color}    
\usepackage{amsmath,amsthm}
\begin{document}

\title{Hidden Complexity in the Isomerization Dynamics of Holliday Junctions}


\author{Changbong Hyeon$^{1*}$, Jinwoo Lee$^2$,  Jeseong Yoon$^1$, Sungchul Hohng$^2$\& D. Thirumalai$^{3*}$}

\thanks{To whom correspondence should be addressed. Email: hyeoncb@kias.re.kr or  thirum@umd.edu}

\affiliation{$^1$School of Computational Sciences, Korea Institute for Advanced Study, Seoul 130-722, Korea
$^2$ Department of Physics and Astronomy, Seoul National University, Seoul 151-747, Korea
$^3$ Institute for Physical Science and Technology, University of Maryland, College Park, MD 20742, USA}

\begin{abstract}
A plausible consequence of rugged energy landscapes of biomolecules is that functionally competent folded states may not be unique, as is generally assumed.
Indeed, molecule-to-molecule variations in the dynamics of enzymes and ribozymes under folding conditions have recently been identified in single molecule experiments. However, systematic quantification and the structural origin of the observed complex behavior remain elusive. 
Even for a relatively simple case of isomerization dynamics in Holliday Junctions (HJs), molecular heterogeneities persist over a long observation time ($\mathcal{T}_{obs}\approx 40$ sec).  
Here, using concepts in glass physics and complementary clustering analysis, we provide a quantitative method to analyze the smFRET data probing the isomerization in HJ dynamics. 
We show that ergodicity of HJ dynamics is effectively broken; as a result, the conformational space of HJs is partitioned into a folding network of kinetically disconnected clusters. While isomerization dynamics in each cluster occurs rapidly as if the associated conformational space is fully sampled, distinct patterns of time series belonging to different clusters do not interconvert on $\mathcal{T}_{obs}$. 
Theory suggests that persistent heterogeneity of HJ dynamics is a consequence of internal multiloops with varying sizes and flexibilities frozen by Mg$^{2+}$ ions. 
An annealing experiment using Mg$^{2+}$-pulse that changes the Mg$^{2+}$ cocentration from high to low to high values lends support to this idea by explicitly showing that interconversions can be driven among trajectories with different patterns. 
\end{abstract}
\maketitle

\section{Introduction}
Although challenged occasionally \cite{Frieden79ARBiochem,Schmid81EJB,Honeycutt90PNAS}, it has long been considered as a general principle that the folded states of biomolecules 
are uniquely determined by their sequences and environmental conditions \cite{Anfinsen75APC}.
Observation of multiple folding paths and the dependence of the folding routes on the initial conditions \cite{Silverman00BC,TreiberCOSB99,Russell02PNAS,Thirum05Biochem} that illustrate the ruggedness of the folding landscapes \cite{XiePNAS04,Bustamante03Science,ZhuangSCI02,Ditzler08NAR,Mickler07PNAS} have illustrated the complexity of biomolecular foldings. These findings by themselves do not challenge the notion of a unique native state.   
However, recent findings from single molecule experiments on several biomolecular systems  explicitly showed persistent heterogeneities in time traces (or molecule-to-molecule variations) generated under identical folding conditions \cite{Frauenfelder1988ARBBC,ZhuangSCI02,Ditzler08NAR,Okumus04BJ,Solomatin10Nature,Borman10CEN}. 
Unlike phenotypic cell-to-cell variability among genetically identical cells that can be visualized using microscope \cite{Pelkmans12Science}, the heterogeneity among individual biomolecules on much small length scales is tantalizing in that it is difficult to reconcile with the conventional notion that functional states of proteins and RNAs are unique or that various native basins of attraction easily interconvert. 
For example, in docking-undocking transitions of surface immobilized hairpin ribozyme \cite{ZhuangSCI02} and \emph{Tetrahymena} group I intron ribozyme \cite{Solomatin10Nature},
each time trace for individual molecule displays very different dynamic pattern with a long memory without apparent compromise in catalytic efficiency; thus it was suggested that these ribozymes have multiple native states \cite{Solomatin10Nature}. 
Control experiments under vesicle encapsulation still displayed heterogeneities as those under surface-immobilization, which indicates that the heterogeneities are intrinsic to molecules being probed, and is not an instrumental artifact \cite{Ditzler08NAR,Okumus04BJ,Rasnik05ACR}.
Given the ubiquity of molecule-to-molecule variations in single molecule experiments, it behooves us to devise analytic tools to quantify the observations using a rigorous theoretical treatments based on statistical mechanics. 
Furthermore, a molecular system that is simpler than the structured RNAs mentioned above but still displays persistent conformational heterogeneity could allow us to glean the molecular origin of the heterogeneity. To this end we study the dynamics of Holliday Junctions that undergo globally a simple two-state like isomerization transition in the presence of Mg$^{2+}$ ions, but reveal a complex behavior when examined in detail.

Time series data from single molecule measurements can reveal the rate of conformational space navigated by a molecule \cite{Flomenbom06PNAS,Komatsuzaki08PNAS}. 
Thus, the manifestation of molecule-to-molecule variation in the measured trajectories, which is often hard to quantify because the time series of an observable in single molecule measurement of biomolecules results from a projection of dynamics in high dimensional space onto a lower dimension, implies that each molecule samples only a small subset of the entire conformational space on $\mathcal{T}_{obs}$. 
The extent to which a trajectory samples the allowed conformational space depends on the length of the observation time $\mathcal{T}_{obs}$.  
Conversely, states hidden in multiple deep furrows of the folding landscapes at high dimension restricts the dynamics of each molecule to one of many states that are non-interconvertible within $\mathcal{T}_{obs}$.

Historically the widely accepted notion of unique native state in biomolecules was hypothesized based on bulk measurements, where an averaged property of a probe variable is obtained from an ensemble of snapshots. Such a conclusion  assumes ergodicity, i.e., the equivalence between time and ensemble average of an observable. 
Although ergodicity is a necessary condition for equilibrium systems, it is in practice difficult to realize because of unlikelihood that in a single time trace a molecule can sample the entire configurational space \cite{MaBook}.  
Furthermore, the situation is further exacerbated because nominally the observation time in practice, or relevant time scales for many biological phenomena are limited. 
In a rugged landscape, a molecule with an initial conformation distinct from others would repeatedly sample distinct region of the folding landscape for a long observation time, which could be longer than ``biologically relevant time scale". 
This scenario results in heterogeneous dynamics, and ensemble averaging would obscure the complexity of the structural features of the underlying landscape. 
Thus, in this sense the ergodicity of the system is ``effectively" broken. 
Now that molecular heterogeneities are clearly demonstrated in many single molecule data for a variety of unrelated systems it is a major theoretical challenge to devise practical tools to reconstruct rugged folding landscapes from time series.

Here, we perform smFRET experiments \cite{Schuler08COSB}, and use concepts from glass physics \cite{ThirumalaiPRA89,Kirk89JPhysA} and complementary clustering algorithms \cite{Tamayo99PNAS,Sturn02Bioinformatics} to make systematic analysis on smFRET data of metal-ion driven conformational changes in a Holliday Junction (HJ) with a DNA sequence that disallows branch migration \cite{Joo04JMB}. 
We show quantitatively that although the HJ dynamics at the ensemble level can be pictured using a two-state model, the ergodicity of the system is effectively broken.  
The associated folding landscape of HJs is visualized in terms of rarely interconverting multiple ordered states embedded in the two isoforms of HJs. 
Furthermore, the simplicity of HJ structure is explored to discover the structural origin of the heterogeneities in dynamics. 
The presence of internal multiloop topologies with varying sizes and flexibilities at the junction and the high local concentration of Mg$^{2+}$ ions around it, the latter of which is calculated using molecular dynamics simulation, lead us to propose that the heterogeneities in HJ isomerization are due to non-interconverting multiloop topologies ``pinned" by the complexed Mg$^{2+}$ ions. 
We validate the structural explanation by showing that trajectories with different patterns interconvert by an annealing protocol involving cycling the Mg$^{2+}$ ions from high to low to high concentrations.

\section{Results and discussion}

\noindent{\bf smFRET measurements of HJ dynamics and analysis using ensemble averaging.} 
In DNA recombination, HJs are essential intermediates for strand exchange (Fig.1a) \cite{Lushnikov03JBC}. 
At Mg$^{2+}$ concentrations exceeding $\sim $ 50 $\mu$M, HJs exist in two distinct isoforms (\emph{iso-I} and \emph{iso-II}) both of which have the characteristic X-shaped architectures. 
According to previous studies \cite{duckett1990EMBOJ,Joo04JMB}, in the absence of divalent ions HJs have stable open square structure whose apparent FRET efficiency value is $E\approx 0.3$, but at [Mg$^{2+}$]$>50$ $\mu$M make conformational  changes between two stable conformers \emph{iso-I} and \emph{iso-II}, via the open square structure.  
We performed smFRET experiments for a range of Mg$^{2+}$ ion concentrations to monitor the Mg$^{2+}$-dependent isomerization dynamics of surface-immobilized HJs by attaching Cy3 and Cy5 dyes to the terminus of X and R branches (Fig. 1a. See Supplementary Text 1). 
The time dependent FRET efficiency $E_i(t)$ for $i$-th molecule ($i=1,2,\ldots N$) was calculated by taking the ratio between the emission signals ($I_{A,i}(t)$ and $I_{D,i}(t)$) from acceptor and donor dyes using $E_i(t)=I_{A,i}(t)/(I_{A,i}(t)+I_{D,i}(t))$ (Fig. S3). 
The individual time trajectories of $E_i(t)$ monitored for $\mathcal{T}_{obs}\approx 40$ sec at [Mg$^{2+}$]$\geq 0.5$ mM shows multiple transitions between high ($E\approx 0.5$) and low FRET ($E\approx 0.2$) values, corresponding to the \emph{iso-I} and \emph{iso-II} conformers, respectively. 
Histograms of FRET values collected over the entire observation time and population, $P_{ens}(E)$,
reveal bimodal distributions for [Mg$^{2+}$]$\geq 0.5$ mM, all of which can nicely be fit to double-Gaussian functions, and unimodal distribution for [Mg$^{2+}$] = 0.0 mM (Fig. 1b and Supplementary Fig. S4).    
With increasing [Mg$^{2+}$], the two peaks on $P_{ens}(E)$ separate more clearly, indicating that the transitions is becoming increasingly cooperative. 
In addition, the ensemble average of FRET efficiency $\langle E\rangle=\int_0^1EP_{ens}(E)dE$ is invariant at [Mg$^{2+}$]$\geq 0.5$ mM (Supplementary Fig. S4), which is in accord with the previous finding \cite{Joo04JMB} that showed that the relative population of the two isoforms is entirely controlled by the junction sequences and not by the ionic conditions. 

In addition to the ensemble-averaged thermodynamic measure $P_{ens}(E)$, transition kinetics,  which occurs on an average time of $\tau_{I\leftrightarrow II}\approx (0.1-1)$ sec, 
between the two isomers provide a glimpse into how HJ isomerizations occur over varying Mg$^{2+}$ conditions. 
As suggested by the shapes of the Mg$^{2+}$ ion-dependent $P_{ens}(E)$ (Supplementary Fig. S4), the transitions between the two conformers slow down with increasing Mg$^{2+}$ ion concentration, presumably due to an increase in the free energy barrier at higher concentration. 
The survival probabilities ($\Sigma(t)$) calculated from the dwell time distributions ($p_{dwell}(t)$) for both high and low FRET values $\Sigma(t)\left[=1-\int_0^tp_{dwell}(\tau)d\tau\right]$ can be approximately fit using a single exponential function although 5 \% of population exhibit a deviation from the fit. 
Thus, it is tempting to surmise that an approximate two-state picture is adequate to describe the transition between the two means \cite{Joo04JMB} (Fig. 1b and Supplementary Fig. S5).  
\\

\noindent{\bf Molecule-to-molecule variation in individual time traces.} 
The two-state model for the Mg$^{2+}$ ion-dependent isomerization dynamics of the HJ, gleaned from averaging over an ensemble of molecule, ignores the intrinsic heterogeneities among the trajectories.   
However, an inspection of a few disparate trajectories makes clear the dramatic variations between individual molecules.  
In Fig. 2a, one can immediately identify the differences between the three exemplary trajectories that maintain the characteristic pattern of dynamics on $\mathcal{T}_{obs}$. 
The multiple transitions in each time trace render the time average of $E_i(t)$, i.e., $\varepsilon_i(t)\left[=\frac{1}{t}\int^{t}_0d\tau E_i(\tau)\right]$ stationary, which suggests that the conformational space associated with that particular trajectory is exhaustively sampled.  
Thus, using such a trajectory, it is legitimate to calculate a stationary distribution $p_s(E;i)$ = $\lim_{t\rightarrow\mathcal{T}_{obs}}{p(E,t;i)}$ \cite{GardinerBook}, which represents the population of microstates probed on $\mathcal{T}_{obs}$ in terms of FRET efficiency value. 
Notably, despite a number of rapid transitions in each time trace, a finding that is nominally associated with canonical ergodic sampling of conformational space, there are qualitative differences between $p_s(E;i)$ and $P_{ens}(E)$, and among $p_s(E;i)$s with $i=1,2,3$, indicating that the ergodicity of HJ dynamics is effectively broken on $\mathcal{T}_{obs}\approx 40$ sec.   

In the literature the overall variation in dynamics and the associated heterogeneities are often conveniently visualized in the form of scatter plot of the mean dwell times at low and high FRET signals for each molecule as in Fig. 2b \cite{Tan03PNAS,Okumus04BJ,Greenfeld11JBC}. 
The mean dwell times for the three time traces in Fig. 2a are encircled in Fig. 2b.  
Note that the scattered data points in Fig. 2b should in principle distribute in a single spot if the HJ dynamics were truly ergodic. 

At an abstract level, a landscape picture implicated from the above findings is the following: 
For a given time trace belonging to a dynamical pattern, corresponding to a specific molecule $\alpha$, $\mathcal{T}_{obs}\approx 40$ sec is long enough to observe multiple isomerization events, so that the time scale for single isomerization between \emph{iso-I} and \emph{iso-II} ($\tau^{\alpha}_{\mathrm{I\leftrightarrow II}}$) is much smaller than $\mathcal{T}_{obs}$ ($\tau^{\alpha}_{\mathrm{I\leftrightarrow II}}\ll \mathcal{T}_{obs}$).  
Thus, HJ explores the conformations in the $\alpha$ state exhaustively; 
however it is not long enough for interconversion to take place between the pattern $\alpha$ and another pattern, say $\beta$, i.e., $\mathcal{T}_{obs}\ll \tau_{conv}^{\alpha\leftrightarrow\beta}$ where $\tau_{conv}^{\alpha\leftrightarrow\beta}$ is the interconversion time between $\alpha$ and $\beta$ states, implying  
that a substantially high kinetic barrier separates the states $\alpha$ and $\beta$. 
Therefore, dynamics of HJs are effectively ergodic within each state on $\mathcal{T}_{obs}$, but $\mathcal{T}_{obs}$ is not long enough to ensure ergodic sampling of the entire conformational space $-$ a situation that is reminiscent of ergodicity breaking in supercooled liquids \cite{ThirumalaiPRA89} (see also Supplementary Text 2).    
\\

\noindent{\bf Assessing ergodicity  from time series.}
In order to demonstrate quantitatively that  interconversion between the multiple states of HJ implicit in smFRET trajectories is unlikely  we use concepts in glass physics \cite{ThirumalaiPRA89, ThirumalaiPRA90}. 
If the entire conformational space of the HJ is effectively sampled during $\mathcal{T}_{obs}$, which would establish ergodicity, then 
the time averaged $E_i(t)$, i.e., $\varepsilon_{i}(t)$ should converge to an ensemble average for all $i$. 
We use a time-dependent metric $\Omega_E(t)=\frac{1}{N}\sum_{i=1}^N\left(\varepsilon_{i}(t)-\overline{\varepsilon(t)}\right)^2$ with $\overline{\varepsilon(t)}\equiv \frac{1}{N}\sum_{i=1}^N\varepsilon_{i}(t)$, introduced to probe the approach to equilibrium in the context  of simulations of supercooled liquids and glasses  \cite{ThirumalaiPRA89,ThirumalaiPRA90}, to analyze smFRET time series data.  
For ``ergodic" systems, $\Omega_E(t)$ converges to zero at $t\rightarrow \infty$. For finite time $t$ with $\Omega_E(t)\neq 0$, it can be shown that $\Omega_E(t)$ decays as $t^{-1}$ asympotically; thus $[\Omega_E(t)/\Omega_E(0)]^{-1}\sim D_Et$ (see Supplementary Text 3 for further details). 
Because the form of the metric $\Omega_E(t)$ is similar to the mean square displacement, the slope of the $[\Omega_E(t)/\Omega_E(0)]^{-1}$ = $D_E$, is interpreted either as an effective diffusion constant or an effective sampling rate of the conformational space projected onto the FRET efficiency coordinate. 

For the entire ensemble of HJ trajectories, however, we find that $[\Omega_E(t)]^{-1}$ is not linear in time, converging to a finite value. The nonlinearity of $[\Omega_E(t)]^{-1}$ indicates that the conformations belonging to distinct states do not mix on $\mathcal{T}_{obs}$, 
which implies  effective ergodicity breaking in the HJ dynamics on $\mathcal{T}_{obs}$ (Fig. 3b) \cite{ThirumalaiPRA89}. Remarkably, this conclusion still holds for HJ trajectories probed on an extended time scale $\mathcal{T}_{obs}\approx 70$ sec (see Supplementary Fig. S6). 
Notably, the variance at $\mathcal{T}_{obs}$, quantified using $\Omega_E(\mathcal{T}_{obs})/\Omega_E(0)$, is greater at higher [Mg$^{2+}$] (Fig. 3b), suggesting that interactions with Mg$^{2+}$ ions are responsible for creating the heterogeneous environment for HJ molecules. 
\\

\noindent{\bf Partitioning the conformational space using complementary clustering algorithm.} 
How many states that do not interconvert on $\mathcal{T}_{obs}$ are needed to fully account for the experimental data? In order to answer this question we 
complement 
the general approach discussed above to quantitatively assess ergodicity from time trajectories with K-means clustering algorithm. This allows us to partition the conformational space of HJ into multiple ``ergodic subspaces".
The K-means clustering algorithm partitions data into K clusters, so that the distance of data belonging to the cluster to the cluster mean is minimized and the inter-cluster distance are maximized. 
In our problem, the data correspond to the stationary distributions of individual time trajectories. 
We partitioned the set of stationary distributions $\{p_s(E;i)|i=1,\cdots,N\}$ into K clusters, ensuring that $[\Omega_E(t)]^{-1}\sim t$ in each cluster (see Supplementary Texts 4, 5 and Fig. S7 for further detail) so that within each cluster the dynamics is ergodic.  
The consequences of our analysis are summarized as follows: 
(i) For [Mg$^{2+}$]=50 mM, $\{p_s(E;i)|i=1,\cdots,N\}$ are partitioned into five clusters (ergodic subspaces) (Fig. 4a. See also Supplementary Text 5 and Fig. S7). 
(ii) The information of data list in each cluster enables us to partition the set of time averaged trajectories ($\{\varepsilon_i(t)\}$) (Fig. 4b) and scatter plot of dwell times (Fig. 4c), shown in Fig. 3a and Fig. 2b, respectively. 
(iii) The effective diffusion constant $D_E$ in $E$-space associated with the conformational sampling of HJs varies widely from one ergodic subspace to another (Fig. 4d). If we assume that the subspace $k=2$, with the largest $D_E$ (Fig. 4d),  has a smooth landscape ($\epsilon_{k=2}=0$), the roughness scale for the subspace $k=4$ is $\epsilon_{k=4}=\sqrt{\log{(0.5/0.07)}}\approx 1.4$ $k_BT$ from $D_E=D_E^oe^{-\beta^2\epsilon^2}$ \cite{ZwanzigPNAS88,Hyeon03PNAS}. 
Comparison between Figs. 4c and 4d shows that the large heterogeneities of dwell times in the scatter plot are mainly due to the molecule belonging to the ergodic subspace with small $D_E$ (e.g. $k=1, 4, 5$). 
The physical criterion, namely the dynamics in each subspace be ergodic,  imposed in our clustering method makes our analysis unique, providing further glimpses into the details of folding landscapes, which are masked in ensemble average quantities $P_{ens}(E)$ and $\Sigma(t)$. 
\\

\noindent{\bf Structural origin of molecular heterogeneity.} 
What is the structural origin of multiple states in HJ, which leads to dynamics heterogeneity?  Simplicity of the HJ structure allows us to infer the structural origin of the multiple states and their roles in complex dynamics.  
First, our calculations of the electrostatic potential and ion distribution using 100 ns molecular dynamics simulations show that Mg$^{2+}$ ions are localized near the junction region and grooves that have high negative charge density (Fig. 5a). 
Second, the results from m-fold algorithm \cite{Zuker03NAR} indicate that the open square form of HJ, representing the secondary structure of HJ that forms transition state (TS)  at the top of the  path connecting the two isoforms at high [Mg$^{2+}$] can have a spectrum of distinct internal multiloop topologies at the junction (Fig. 5b). 
In the absence of branch migration, which is ruled out in our experiments, 
it is conceivable that the topology of the internal multiloop with varying sizes and flexibilities determines both the rate of HJ isomerization and inter-dye distance.   
Variations in isomerization rate and inter-dye distance are reflected in $\{p_s(E;i)\}$ that are partitioned into five clusters. 
Taking these results together, we argue that the secondary structure of HJ especially at the internal multiloop, which mediate the conformational transition between the two isoforms, is pinned by Mg$^{2+}$ ions. Consequently,  the structural rearrangement needed for interconversion between two distinct states within a given isoform is prevented.  
Note that the TS ensemble is quantized and that the actual free energy gaps in the spectrum of transition states in the presence of Mg$^{2+}$ ions could be larger than the calculated values because the m-fold algorithm does not include the effect of the specific binding of multivalent ions (Control experiments show that monovalent (Na$^+$) ions with high concentrations do not lead to bimodal isomerization. See Supplementary Fig.S9). 
In principle, structural rearrangement could occur via transitions from high to low free energy QTSs (Fig. 5b) between the quantized TSs (QTSs), thus allowing for interconversion between the five states. However, the QTSs are only transiently populated during the isomerization process because  transition path times are considerably shorter than the time to cross free energy barriers \cite{Chung2012science}.  
In other words, the life times of the QTSs are very short, so that the rearrangement of local bubbles at the transition state at high Mg$^{2+}$ concentrations is unlikely. The transient nature of the QTSs implies that during multiple rounds of isomerization the local bubble structures are quenched.
In addition to Fig.3b that shows greater molecular heterogeneity at high Mg$^{2+}$ condition, 
our idea of non-interconverting QTS in the presence of Mg$^{2+}$ ions finds support in the previous studies \cite{Lushnikov03JBC,Panyutin94PNAS}, which showed that the obligatory open square form intermediates  slow down DNA branch migration by $\sim$1,000 fold in the presence of Mg$^{2+}$ ions because this process requires the rupture and formation of base pairs. 

The calculations, summarized in Fig. 5, explain our experimental findings. 
In the  folding landscape of HJ emerging from our analyses (Fig. 5c), transitions are only allowed between \emph{iso-I} and \emph{iso-II} via 
``a band of QTSs'' within which the free energy gap is small enough to allow interconversion on $\mathcal{T}_{obs}$.  
Lack of transitions between two different states (say $\alpha$ and $\beta$) within a given isoform ($\mathcal{T}_{obs}\ll\tau^{\alpha\leftrightarrow \beta}_{conv}$) is explained by noting that rupture of Mg$^{2+}$-stabilized base  pairs are required for rearrangements from one multiloop topology to another. 
The conformational space connecting \emph{iso-I} and \emph{iso-II} is partitioned into a number of kinetically disjoint states ($\xi=\alpha,\beta,\gamma \ldots$), reflecting the band structure of the QTS ensemble.  
In this sense the persistent pattern of a smFRET trajectory is an imprint of specific disjoint states in the rugged folding landscape. \\

\noindent{\bf Mg$^{2+}$ pulse annealing experiments.} An immediate prediction of our model (Fig. 5c) is that interconversion between states $\alpha$ and $\beta$ should be facilitated by an annealing protocol, enabling the release of Mg$^{2+}$ ions from frozen internal multiloop structures.   
In order to validate this prediction 
we performed  single molecule experiments using  a Mg$^{2+}$ pulse sequence [Mg$^{2+}$] = 50 mM $\rightarrow$ 0 mM $\rightarrow$ 50 mM to induce transition between multiple states (Fig. S2).
The annealing experiments confirmed that washing the Mg$^{2+}$ ions from HJ molecules indeed 
facilitates interconversion between trajectories with distinct patterns (compare the trajectories or two $p_s(E;i)$s shown on the side of each panel in Fig. 6a calculated from the blue and red intervals of the trajectories, corresponding to the moment before and after the Mg$^{2+}$ pulse). 
We also calculated the Euclidean distance of $p_s(E;i)$ to the centroid of the five clusters in Fig. 4a before and after the Mg$^{2+}$ pulse annealing experiments. Here, the centroid is the arithmetic mean of $\{p_s(E;i)|i\in k\}$ with $k=1,2,\cdots 5$ (Fig.4a). 
The distances of $p_s(E;i)$ to the cluster means change after the Mg$^{2+}$ pulse, which is also reflected in the reshuffling of the population of the HJ molecules among the five clusters.
Consequently, 
transitions that are prohibited on the time scale of 40 sec are induced as seen by a redistribution of the population among the five distinct clusters (Fig. 6b).  
Resetting the initial memory of each HJ molecule is achieved by temporarily removing the Mg$^{2+}$ from the solutions, enabling the conformational interconversions in the otherwise non-interconverting HJ time traces.  
The observation of facilitated interconversion using Mg$^{2+}$ pulse corroborates the idea that Mg$^{2+}$ ions and their interaction with HJ structure are the major cause of the persistent conformational heterogenetiy in HJ, ruling out the possible contributions to the molecular heterogeneity from experimental artifacts such as aforementioned surface immobilization \cite{Okumus04BJ}, heterogeneous dye stacking with DNA bases \cite{Iqbal08PNAS}, or chemical modifications \cite{Greenfeld11JBC} (Supplementary Fig. S10 shows the non-responsiveness of  donor-only tagged HJs to the Mg$^{2+}$ pulse, corroborating that the Mg$^{2+}$ ions affect the internal conformational dynamics, not the dye stacking or dye-to-surface interaction). 
To recapitulate, in HJs, Mg$^{2+}$ ions play a key role not only in slowing down the isomerization by stabilizing the native states in two isoforms (Supplementary Figs. S4, S5) but also in reducing the probability of interconversion between two different HJ trajectories by enlarging the free energy gap of the QTS ensemble (Fig. 5b).
\\

\section{Conclusions}
Single molecule measurements provide new avenues to probe the dynamics of biological systems by providing information not available in ensemble experiments. However,  
most of the current experimental studies build distribution of observable or make dwell time analysis by ensemble averaging, thus overlooking molecular heterogeneity. 
Our novel theoretical analysis employing the concept of ergodicity breaking provides a practical framework to analyze single molecule data and to decipher complex folding landscapes of biological systems. 
Only by quantitatively analyzing each trajectory individually, without succumbing to the temptation to  average, the dynamical complexity of biological molecules can be fully revealed. 
Indeed, as noted recently, quantifying and understanding the consequence of non-ergodic behavior of RNA molecules is a major challenge \cite{AlHashimi08COSB}, and the present work provides the  framework for meeting it.  Finally, it is worth noting that although we have used HJ as example to explore quantitatively the concept of heterogeneity  our conclusions are far reaching.  We expect similar behavior in biological systems spanning spatial scales from nm to several microns (cell dimensions) and time scales from $\mu$s to minutes and longer. 
\\



This work was supported in part by the grants from National Research Foundation of Korea (2010-0000602) (to C.H.), the Creative Research Initiatives (Physical Genetics Laboratory, 2009-0081562) (to S.H.) and National Science Foundation Grant CHE 09-14033 (to D.T.).   
\\
 
 J.L. and S.H. carried out smFRET measurements on HJs under varying Mg$^{2+}$ concentrations and Mg$^{2+}$ pulse. 
 C.H. carried out the smFRET data analysis.  
 J.Y. carried out all-atom molecular dynamics simulations to determine the radial distribution of Mg$^{2+}$ ions around HJs. 
 C.H. and D.T. conceived and directed the project, and prepared the manuscript.  
\\









\clearpage
\begin{figure}
\includegraphics[width=6.2in]{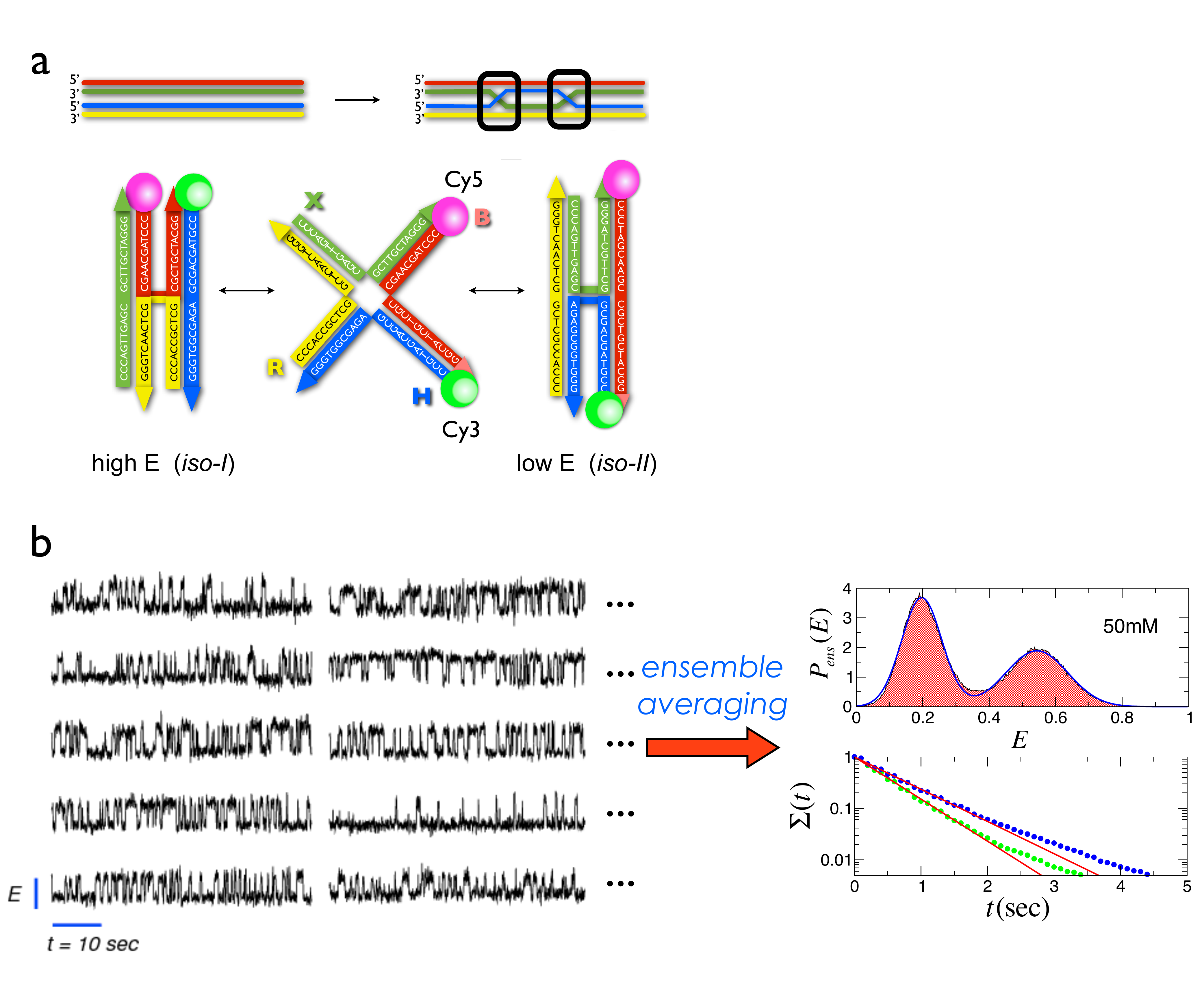}
\caption{HJ dynamics probed with smFRET experiments and their analysis using conventional ensemble averaging method. 
{\bf a.} A schematic of strand exchange in DNA recombination (top) and the two isoforms connected by the open square structure (bottom). The Cy5 (magenta) and Cy3 (green) dyes attached to the B and H branches for the smFRET measurement are depicted as spheres.  
{\bf b.} Shown are the part of FRET time traces ($\{E_i(t)\}$ with $i=1,2,\ldots,N$ with $N=315$) obtained for individual HJ molecules at [Mg$^{2+}$] = 50 mM. Similar to the findings in the literatures, the conventional analysis using the ensemble of these time traces without deliberating the molecule-to-molecule variation gives rise to an interpretation of the apparent two-state behavior for HJs. The ensemble averaged histogram of the FRET efficiency $E$, i.e., $P_{ens}(E)$, is  nicely fit to a double-Gaussian curve (blue line), and the dwell time distribution (bottom panel) for low (data in green) and high (data in blue) FRET states are approximately fit to single exponential functions (red lines).}
\end{figure} 
\clearpage

\begin{figure}
\includegraphics[width=6.2in]{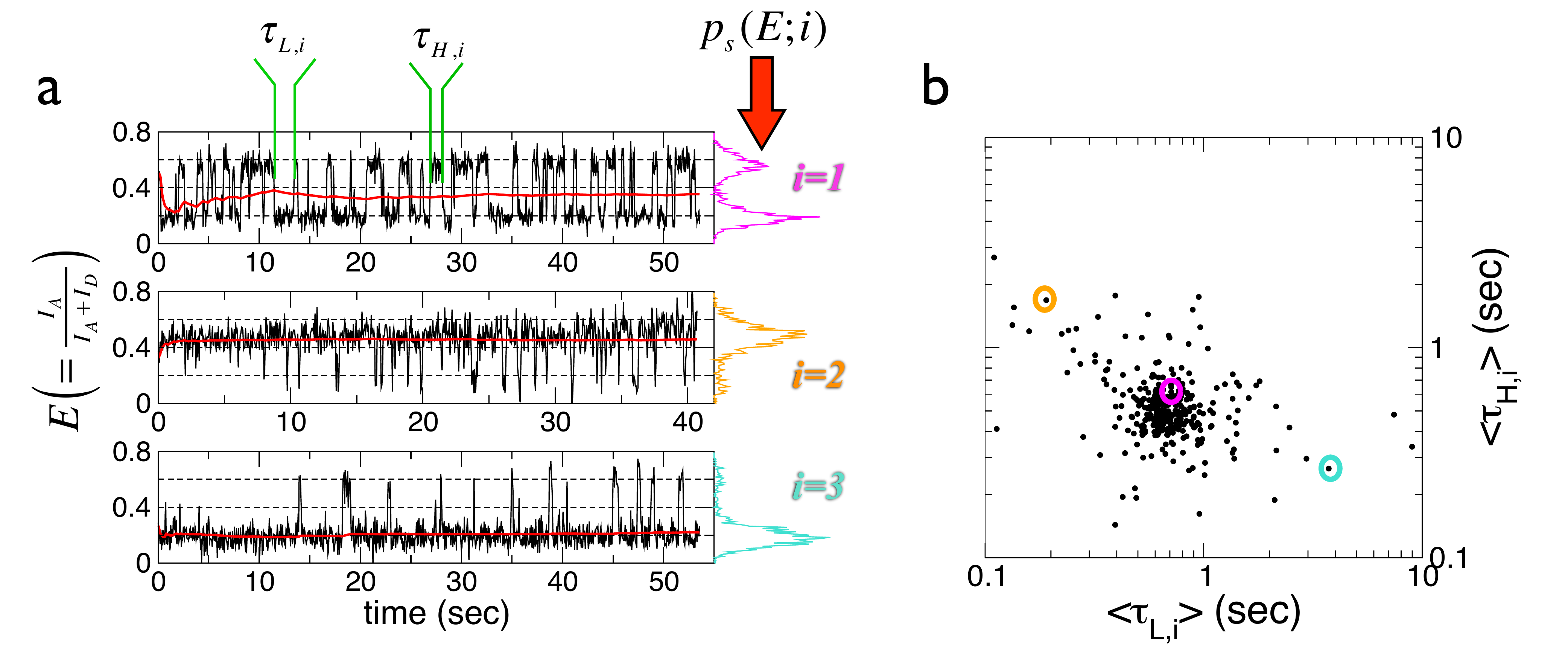}
\caption{Molecule-to-molecule variation (or molecular heterogeneity) manifested in the time traces of isomerization dynamics of HJs
{\bf a.} Three disparate FRET time traces $E_i(t)$ at [Mg$^{2+}$]=50 mM (blue), their time average $\varepsilon_i(t)$ (red), and the corresponding histograms on the right. 
Because each time trace can be considered stationary as clearly indicated in the time-independence of $\varepsilon_i(t)$, it is legitimate to build a histogram for each time trace and designate the histogram a stationary distribution, $p_s(E;,i)$. 
{\bf b.} Molecular heterogeneity revealed in the scatter plot of average dwell times at low and high FRET state for individual time traces ($\langle \tau_{L,i}\rangle$, $\langle \tau_{H,i}\rangle$). The encircled data points correspond to the three time traces in {\bf a}.}
\end{figure}
\clearpage

\begin{figure}
\includegraphics[width=6.2in]{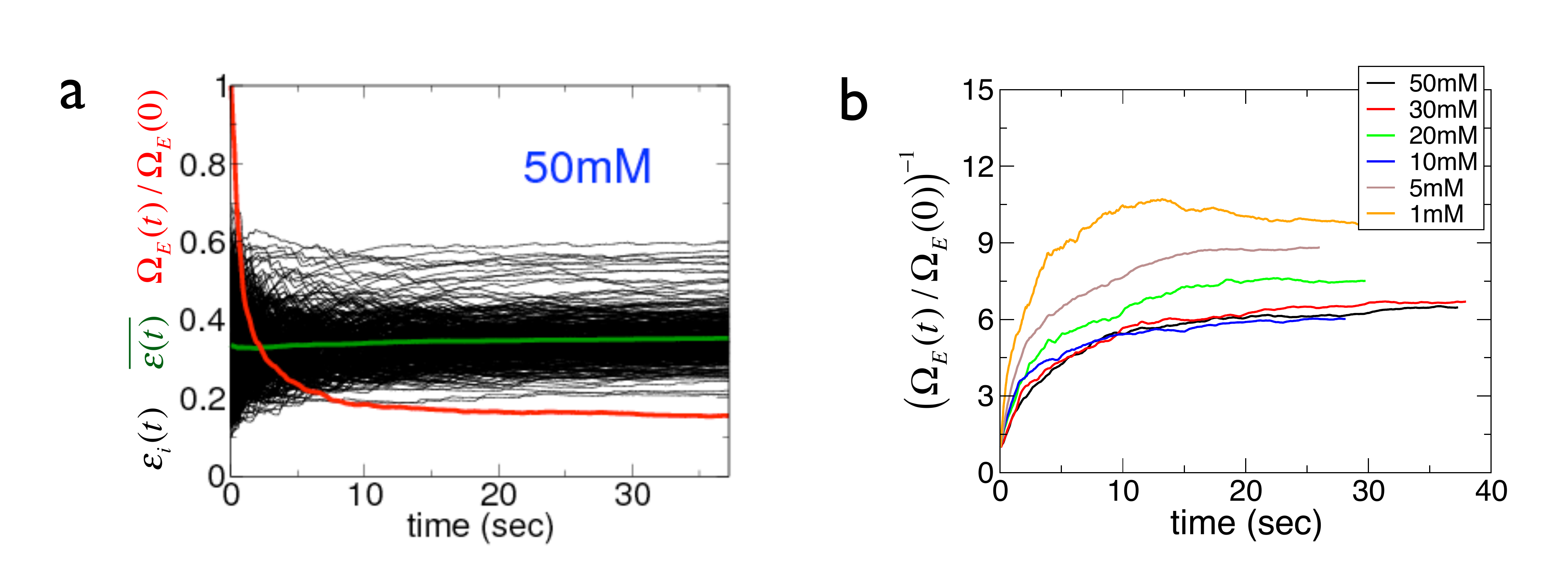}
\caption{Probing the ergodicity breaking.
{\bf a}. Probes of ergodic behavior in smFRET trajectories using the metric $\Omega_E(t)$ at [Mg$^{2+}$]=50 mM. 
$\varepsilon_{i}(t)$, $\overline{\varepsilon(t)}$,  and $\Omega_E(t)/\Omega_E(0)$, (defined in the text) are shown in black, green, and red lines, respectively.
{\bf b}. The nonlinearity of $[\Omega_E(t)]^{-1}$ for [Mg$^{2+}$]$>1$ mM shows that the HJ dynamics is non-ergodic for all concentrations on the observation time $T_{obs}$. 
}
\end{figure}
\clearpage

\begin{figure}
\includegraphics[width=3.2in]{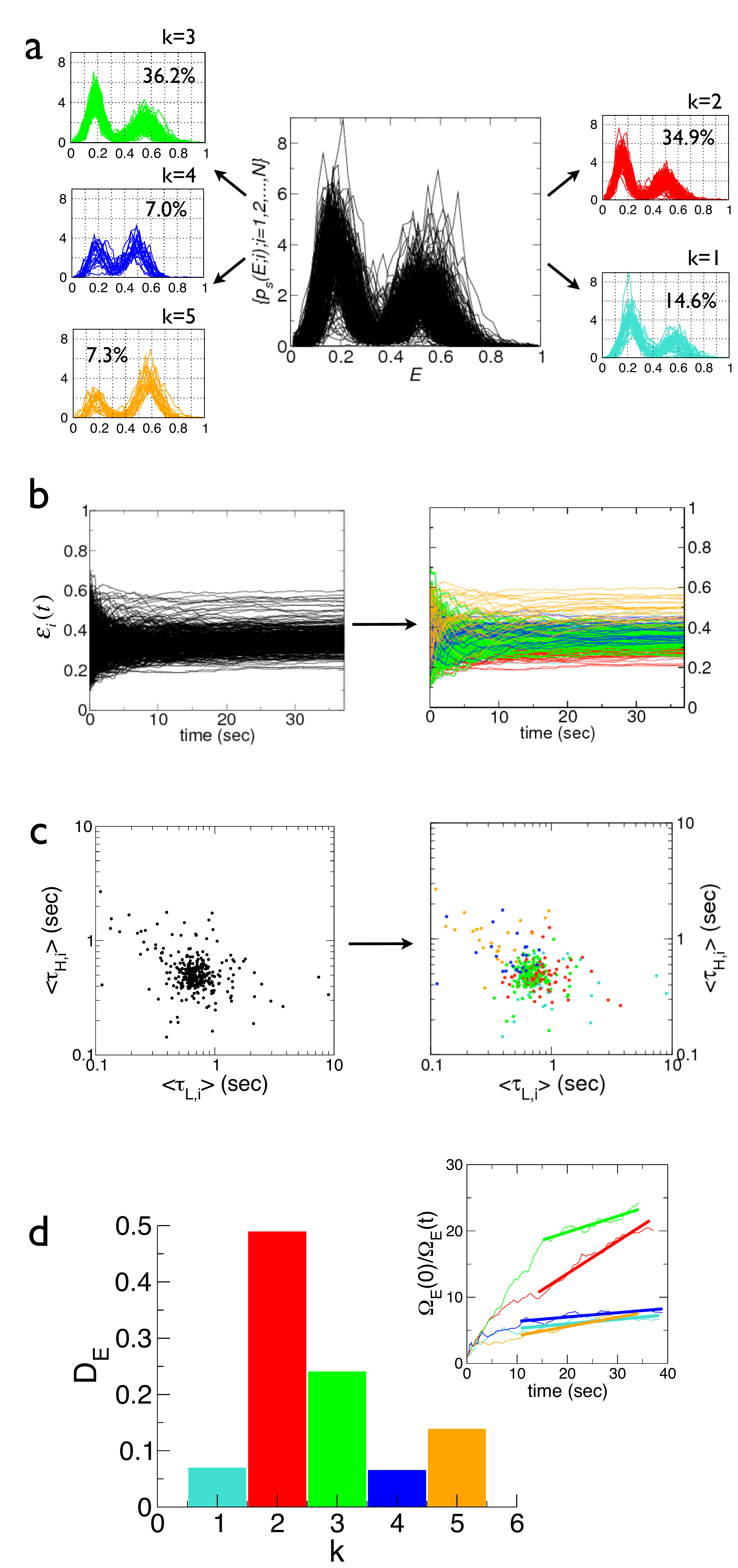}
\caption{Partitioning the molecules into distinct clusters.  
{\bf a}. K-means clustering algorithm combined with ergodic criteria partitions the set of stationary distributions $\{p_s(E;i)\}$ into 5 clusters for [Mg$^{2+}$]= 50 mM, and determines the list of time traces that belongs to the clusters from $k=1$ to 5.
{\bf b}. The list of time traces for each cluster determined in a is used to partition 
$\{\varepsilon_i(t)\}$ into  $\{\varepsilon_i(t)|i\in k\}$ for $k=1,2,\cdots 5$. 
{\bf c}. $D_E$ value are calculated from the fits using $\Omega_E(0)/\Omega_E(t)\sim D_Et$ for each cluster ($k=1,\ldots,5$). 
{\bf d}. Clustering of dwell time data as a result of the $\{p_s(E;i)\}$ clustering.
}
\end{figure}
\clearpage

\begin{figure}
\includegraphics[width=6.0in]{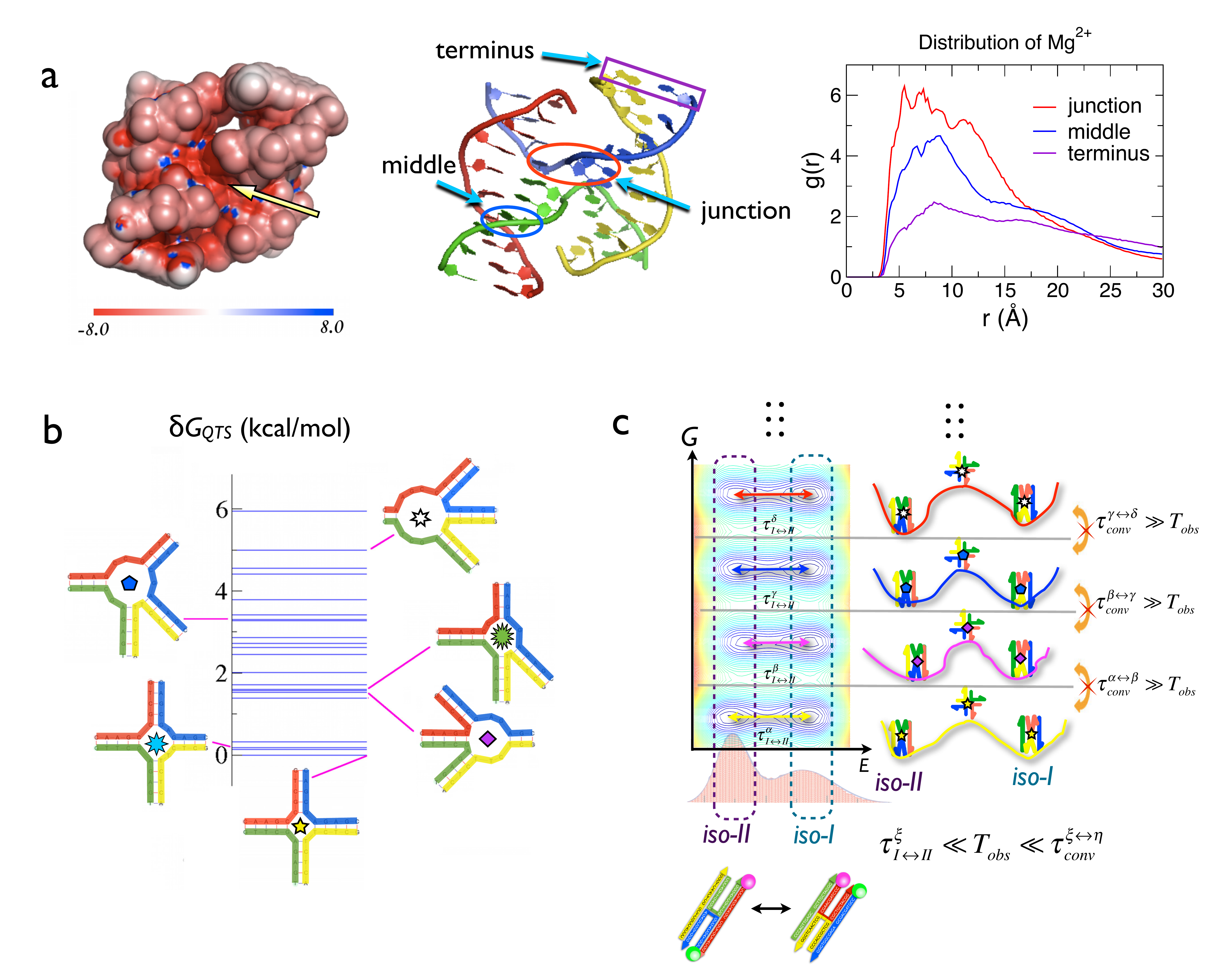}
\caption{Structural model to account for the origin of molecule-to-molecule variation in the HJ dynamics. 
{\bf a.} (left) Electrostatic potential calculated at 200 mM monovalent ion condition using an X-ray structure of HJ (PDB code: 1DCW) (see Supplementary Text 6).
The energy scale for the potential is in $k_BT/e$ unit.
(middle, right) 100 $nsec$ MD simulation at $T=310$ K (Supplementary Text 7) shows that Mg$^{2+}$ ions are localized more at the junction and grooves than at the terminus.  
{\bf b.} HJs with various topologies of internal multiloops, which are the putative QTSs (quantized transition states) connecting one state in \emph{iso-I} and another in \emph{iso-II}. 
{\bf c.} Model for the dynamics of HJ constructed based on smFRET experiments and simulations. 
On the left are the free energy contours for various states. 
Two isoforms in each state are connected by a distinct open square form, whose structures are shown in B.  Ensemble averaged distribution of the FRET efficiencies, $P_{ens}(E)$, is shown at the bottom. 
On the right schematic of the free energy profiles are shown with the cartoons of HJ structures; the symbols (star, pentagon, $\ldots$) at the junction emphasize that the junction structure is intact during the isomerization process.  
Hence,  $\tau_{\mathrm{I\leftrightarrow II}}^{\xi}$ ($\xi = \alpha, \beta\ldots$) $\ll\mathcal{T}_{obs}\ll\tau_{conv}^{\xi\leftrightarrow\eta}$ ($\xi,\eta=\alpha,\beta,\gamma,\ldots$ with $\xi\neq\eta$) is established. 
}
\end{figure}
\clearpage 

\begin{figure}
\includegraphics[width=5.2in]{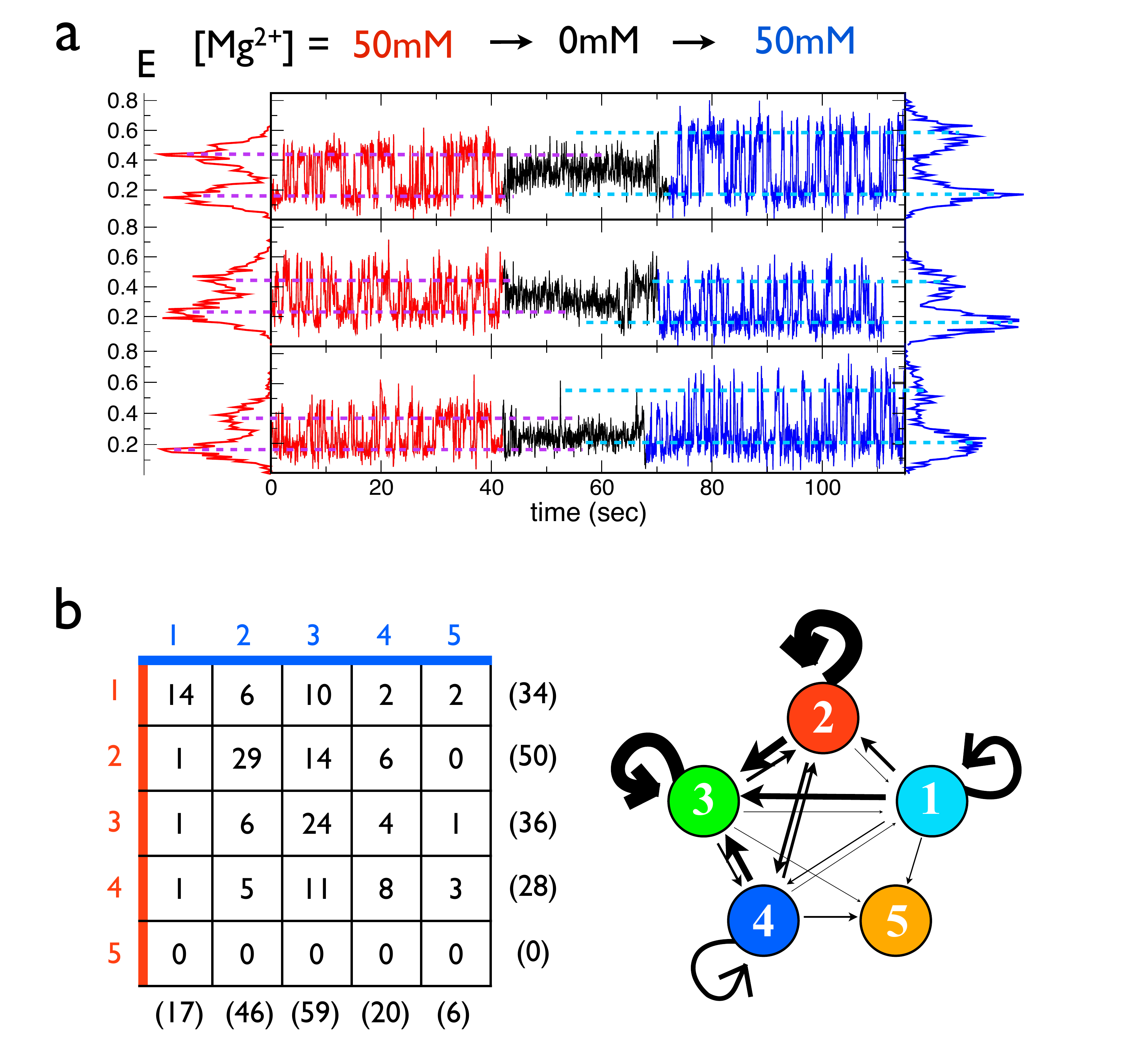}
\caption{Mg$^{2+}$ pulse experiments to reset the molecular population in conformational space.  
{\bf a.} Effect of Mg$^{2+}$ pulse experiment. Shown are three representative trajectories that change their pattern ($E(t)$ or $p_s(E;i)$) in response to Mg$^{2+}$ pulse sequence. 
The dashed lines depict the peak positions of $p_s(E;i)$ and underlies the differences in $p_s(E;i)$ before (red) and after (blue) washing off Mg$^{2+}$ ions. 
{\bf b.} 
Depicted are Mg$^{2+}$ pulse-induced transition frequency matrix and diagram among kinetically disjoint states based on 148 FRET trajectories. 
The indices at the sides of matrix and in the nodes denote the cluster number $k=1,2\ldots 5$. 
The numbers in the parentheses are the occupation number in each cluster, which can be obtained by summing up the transition frequency from one cluster to the other.
The diagram on the right is the kinetic network describing the HJ transition under Mg$^{2+}$ pulse.  The widths of the arrows are in proportion to the number of transitions. 
}
\end{figure}
\clearpage 

\section*{Supplementary Information}
\begin{enumerate}
\item Single molecule FRET experiments $\ldots\ldots\ldots\ldots\ldots\ldots\ldots\ldots\ldots\ldots\ldots\ldots\ldots\ldots\ldots$ 22
\item Memory of initial condition $. \ldots\ldots\ldots\ldots\ldots\ldots\ldots\ldots\ldots\ldots\ldots\ldots\ldots\ldots\ldots\ldots\ldots$ 23
\item Probing the ergodicity breaking$.\ldots\ldots\ldots\ldots\ldots\ldots\ldots\ldots\ldots\ldots\ldots\ldots\ldots\ldots\ldots\ldots$ 24
\item Clustering the stationary FRET distribution functions $.\ldots\ldots\ldots\ldots\ldots\ldots\ldots\ldots\ldots .$  25
\item Clustering at the optimal values of K $.\ldots\ldots\ldots\ldots\ldots\ldots\ldots\ldots\ldots\ldots\ldots\ldots\ldots\ldots .$ 26
\item Calculation of electrostatic potential maps of HJ structures $\ldots\ldots\ldots\ldots\ldots\ldots\ldots\ldots$  26
\item Calculation of radial distribution function of Mg$^{2+}$ ions around HJ $. \ldots\ldots\ldots\ldots\ldots .$  26
\end{enumerate}
\begin{itemize}
\item References and Notes $. \ldots\ldots\ldots\ldots\ldots\ldots\ldots\ldots\ldots\ldots\ldots\ldots\ldots\ldots\ldots\ldots\ldots\ldots\ldots$ 27
\item Figures S1 - S10 $\ldots\ldots\ldots\ldots\ldots\ldots\ldots\ldots\ldots\ldots\ldots\ldots\ldots\ldots\ldots\ldots\ldots\ldots\ldots\ldots$ 28-37
\end{itemize}
\clearpage 

\setcounter{section}{0}
\renewcommand*\thesection{\arabic{section}}
\section{Single molecule FRET experiments}
To assemble the Holliday junction, the following DNA sequences were purchased from IDTDNA (Coralville, IA). 

R branch:    5'-biotin-TTTTTTTT {\it CCCACCGCTCG}$\underbrace{\mathrm{GCTCAACTGGG}}$-3'

H branch:                           5'-Cy3-\underline{\underline{CCGTAGCAGCG}}{\it CGAGCGGTGGG}-3'

X branch: 5'-GGGCGGCGACCT $\underbrace{\mathrm{CCCAGTTGAGC}}${\bf GCTTGCTAGGG}-3'

B branch:                           5'-Cy5-{\bf CCCTAGCAAGC}\underline{\underline{CGCTGCTACGG}}-3'





\noindent where the sequences with identical font style represent the pair of complementary DNA strands. The DNA strands were annealed by heating the mixture of DNAs (1 $\mu M$, 20 $\mu l$) to 90 $^o$C in TN buffer (10 mM Tris with 50 mM NaCl, pH 8.0), and slowly cooling down to 4 $^o$C with 1 $^o$C/min cooling rate. 
For single-molecule FRET (smFRET) measurements, a sample chamber was made between a cleaned quartz microscope slide (Finkenbeiner) and a cover slip using double-sided adhesive tape. DNAs were immobilized on quartz surface by successive additions of biotinylated BSA (40 $\mu l$, 1 mg/ml, Sigma-Aldrich), streptavidin (40 $\mu l$, 0.2 mg/ml, Invitrogen), and DNA in TN buffer. 
The concentration of DNA was adjusted to achieve proper single molecule density. 
Each injection was incubated for 2 minutes, which was followed by washing with TN buffer. 
Experiments were performed at room temperature in a conventional imaging buffer (10 mM Tris-HCl with 0.4\% (w/v) glucose (Sigma-aldrich), 1\% (v/v) trolox (Sigma-aldrich), 1mg/ml glucose oxidase (Sigma-aldrich), 0.04 mg/ml catalase (Roche, Nutley, NJ), and designated magnesium ion) by using a home-built prism-type total internal reflection single-molecule FRET setup (Fig. S1). 
Specifically, a green laser (532 nm, Compase 215 M, Coherent) was used as an excitation source. 
Fluorescence signals of donor and acceptor were collected through a water-immersion objective (Olympus, UPlanoSApo 60x/1.2w), divided using dichroic mirror (Chroma, 635 dcxr) as a wavelength, and finally focused on different areas of an electron multiplier charge-coupled-device camera (Andor, Ixon DV897ECS-BV). 
To reliably detect individual transitions of HJ, optimum exposure time of the camera was selected at each Mg$^{2+}$ concentration (30 ms for 5 mM and 10 mM Mg$^{2+}$ concentrations, and 50 ms for other concentrations). 
Data acquisition and selection of single molecule traces were done by using home-made programs written in VC++6.0, and IDL (ITT), respectively. 

For Mg$^{2+}$ pulse experiment (Fig. S2), a teflon tube was connected to the exit hole of a sample chamber filled with 50 mM Mg$^{2+}$ buffer, and a pipette tip filled with Mg$^{2+}$-free imaging buffer was carefully plugged into the entrance hole. 
After starting a data acquisition, the buffer solution in the detection chamber was rapidly replaced with a Mg$^{2+}$-free buffer solution in the pipette tip by pulling a syringe tube connected to the Teflon tube. 
Then, remaining solution in the pipette tip was carefully replaced with a 50 mM Mg$^{2+}$ buffer, and the new solution containing 50 mM Mg$^{2+}$ was rapidly introduced into the detection chamber by suction. 
During the whole process, data acquisition was maintained.  
The measured smFRET trajectories, under a range of counterion (Mg$^{2+}$) concentrations, were analyzed using concepts in glass physics and bioinformatic tools. \\

\section{Memory of initial condition}
Another evidence for the inadequacy of using $\Sigma(t)$ and $P_{ens}(t)$ calculated over ensemble in unraveling the rugged folding landscapes can be given by comparing the patterns of FRET trajectories and the probability of observing such trajectories based on two-state picture.  
A series of dwell times of a trajectory generated from a system that precisely exhibits two-state kinetics should obey a renewal (more precisely a Poisson) process with dwell time distribution $p_{dwell}(t)=\tau^{-1}e^{-t/\tau}$, where $\tau$ is the mean dwell time. 
For such a system, the probability of observing a time trace with successively long multiple ($n$) dwells ($\tau^*\gg\tau$), i.e., $\left[\int^{\infty}_{\tau^*}dtp_{dwell}(t)\right]^n=\exp{\left(-n\tau^*/\tau\right)}$ is essentially zero (For $n = 10$ and $\tau^*=5\tau$, the probability of observing such a time trace is less than $10^{-22}$). 
Nevertheless, our smFRET trajectories contain a preponderance of such cases, in which successively long multiple dwells are observed (see Fig. 2a). 
Thus, even during frequent isomerization, a molecule behaves as if it retains ``memory" of its basin of attraction \cite{ZhuangSCI02}, which implies that a given molecule repeatedly visits exactly the same states in \emph{iso-I} and \emph{iso-II}. 
\\

\section{Probing the ergodicity breaking}
For an ensemble of smFRET trajectories that reports the FRET efficiency ($E$) as a function of time ($t$), the fluctuation metric ($\Omega_E(t)$) is defined  \cite{ThirumalaiPRA89}:
 \begin{equation}
\Omega_E(t)=\frac{1}{N}\sum_{i=1}^N\left(\varepsilon_i(t)-\overline{\varepsilon(t)}\right)^2
\end{equation}
where $\varepsilon_i(t)\equiv\frac{1}{t}\int^t_0dsE_i(s)$  is the time average of $E$ for  molecule $i$, and $\overline{\varepsilon(t)}\equiv\frac{1}{N}\sum_{i=1}^N\varepsilon_i(t)$ is the ensemble average of $\varepsilon_i(t)$.   In accord with the ergodic hypothesis we expect that, $\varepsilon_i(t\rightarrow \infty)\equiv\lim_{t\rightarrow\infty}\frac{1}{t}\int^t_0dsE_i(s)=\frac{1}{N}\sum_{i=1}^NE_i(t)=\langle E\rangle$, which leads to $\lim_{t\rightarrow\infty}\overline{\varepsilon(t)}\equiv\frac{1}{N}\sum^N_{i=1}\varepsilon_i(t\rightarrow\infty)=\langle E\rangle$ . Therefore,  the necessary condition, $\lim_{t\rightarrow\infty}\Omega_E(t)\rightarrow 0$, 
should be fulfilled for ergodic systems. Eq.(1) for $\Omega_E(t)$ can be  rewritten as
\begin{equation}
\Omega_E(t)=\frac{1}{t^2}\int^t_0ds_1\int^t_0ds_2\frac{1}{N}\sum_{i=1}^N\left(E_i(s_1)-\langle E\rangle\right)\left(E_i(s_2)-\langle E\rangle\right)=\frac{1}{t^2}\int^t_0ds_1\int^t_0ds_2C(s_1,s_2)
\end{equation}
where $C(s_1,s_2)$ is the equilibrium time correlation function. If ($E_i(t)-\langle E\rangle$) is self-averaging, one can put $C(s_1,s_2)=C(|s_2-s_1|)$. 
Since in equilibrium there is no preferred origin of time $C(s_1,s_2)$ depends only on the difference between the two times $s_1$ and $s_2$. 
With straightforward algebra $\int^t_0ds_1\int^t_0ds_2C(|s_2-s_1|)=2\int^t_0ds_1\int^t_{s_1}ds_2C(s_2-s_1)=2\int^t_0d\tau(t-\tau)C(\tau)$ one gets 
$\Omega_E(t)=\frac{2}{t}\int^t_0d\tau\left(1-\frac{\tau}{t}\right)C(\tau)\rightarrow\frac{2}{t}\int^t_0d\tau C(\tau)$ for asymptotic limit of $t$. Therefore at large $t$ the proposed metric behaves as 
\begin{equation}
\frac{\Omega_E(t)}{\Omega_E(0)}\approx \frac{1}{D_Et}
\end{equation}
where $D_E=\lim_{t\rightarrow\infty}{\left[2\int^t_0ds(C(s)/C(0)\right]^{-1}}$. 
Note that for ergodic systems, the equilibrium time correlation function $C(s)/C(0)$ is a fast decaying function; thus the integral $2\int^{\infty}_0\frac{C(s)}{C(0)}ds$ yields a constant.  
Next, the slope of the inverse of $\Omega_E(t)/\Omega_E(0)$ yields  effective diffusion coefficient, $D_E$, the rate at which the molecule navigates the accessible conformational space.
However, if the erogodicity of the system is effectively broken, $[\Omega_E(t)/\Omega_E(0)]^{-1}$ is no longer linear in time, but saturates into a constant value, which we observed for the ensemble of time trajectories that probe the isomerization dynamics of HJs.    
\\

\section{Clustering the stationary FRET distribution functions} 
The set of stationary distributions calculated for the individual time traces of HJ isomerization, 
$\{p_s(E;i)|i=1,2,\cdots,N\}$, can be partitioned into the K-clusters,
$S(E)$=$(S_1(E),S_2(E),$$\ldots,$$S_K(E))$  with $K < N$, 
by employing K-means clustering algorithm that is widely used as a bioinformatics tool in analyzing the patterns of gene expression \cite{Tamayo99PNAS,Sturn02Bioinformatics}. 
We treated each stationary distribution ($p_s(E;i)$) as a vector in the $E$-space and iteratively minimized the sum of squared Euclidean distance 
from each $p_s(E;i)$ to its cluster centroid, i.e.,   
$\sum_{k=1}^K\sum_{p_s(E;i)\in S_k(E)}||p_s(E;i)-\mu_k(E)||^2$
where $\mu_k(E)$ is the mean of $S_k(E)$. 
After the random assignment of the cluster centroid to a vector in $E$-space,  the K-means procedure iteratively refined partitioning of the $p_s(E;i)$ among the K clusters by assigning each vector to the nearest cluster centroids, then recalculating cluster centroids if cluster membership ($S(E)$) was changed from the previous iteration. 
If no change in any cluster membership occurred then the K-means algorithm was terminated. 
To avoid local minima, the optimization was initiated from 200 random cluster centroids. We selected the best set of clusters that reaches the minimum value among the 200 trials. 

The quality of clustering result can be assessed by using a silhouette function: 
\begin{equation}
s(i)=\frac{b(i)-a(i)}{\max\{a(i),b(i)\}}, 
\end{equation}
where $a(i)$ is the average distance (dissimilarity) of a datum $i$ with all other data within the same cluster, 
and $b(i)$ is the lowest average dissimilarity of the datum $i$ with the data belonging to other clusters.  
The optimal K value is determined by examining the mean silhouette function $\langle s\rangle=\frac{1}{N}\sum_{i=1}^Ns(i)$ against K. 
\\

\section{Clustering at the optimal values of K}
At [Mg$^{2+}$]=50 mM, $\langle s\rangle$ becomes optimal at K=3, 5, and 13. 
However, the criteria using Euclidean distance or Pearson's correlation do not guarantee the ergodicity of each cluster. To this end we evaluated the fluctuation metric $\Omega_E(0)/\Omega_E(t)$ for each cluster to ensure its ergodicity (Fig.~S7). Although maximal $\langle s\rangle$ was found at K=3 (Fig.~S7A), $\Omega_E(0)/\Omega_E(t)$ of the cluster k=2 (red), which represents the 36.5 \% of the time traces, was not linear in $t$ (Fig.~S7B).  
Consequently, we had to inspect the next optimal values of K (K=5 and 13).  
\\

\section{Calculation of electrostatic potential maps of HJ structures}
Electrostatic potential maps of the three Holliday Junctions were calculated by solving the nonlinear Poisson-Boltzmann equation using the APBS software \cite{Baker01PNAS} at 200 mM monovalent ion condition with the dielectric constants for nucleic acids and solvent being 2.0 and 78.5, respectively. The calculated electrostatic potentials were visualized (Fig. 5a in the Main text) using the PyMol program (http://www.pymol.org). 
\\

\section{Calculation of radial distribution function of Mg$^{2+}$ ions around HJ}
To calculate radial distribution function of Mg$^{2+}$ around HJ, we performed all-atom molecular dynamics simulations using the crystal structure of HJ with PDB ID 1DCW. After solvating the HJ molecue in a  66${\rm \AA}$ $\times$ 66${\rm \AA}$ $\times$66${\rm \AA}$ box containing 8369 TIP3P water molecules, 37 Na$^{+}$, 41 Cl$^{-}$, and 20 Mg$^{2+}$ ions were randomly placed. 
The system was minimized by using 1 fs time step for 2000 steps with constraints on DNA coordinates, then for an additional 3000 steps without constraints. 
During equilibration stage we gradually heated the system from 0 K to 310 K using 2 fs time step for 620 ps with constraint, and equilibrated for an additional 5 ns by using the NPT ensemble at 310 K and 1 atm. 
After equilibration, we generated a 100 ns trajectory in the NPT ensemble at 310 K and 1 atm. The integration step during production run is 2 fs. 
The simulations were performed by using NAMD with CHARMM force field. 
\\


\clearpage

\renewcommand{\thefigure}{S1}
\begin{figure}[tbp]
\includegraphics[width=7.0in]{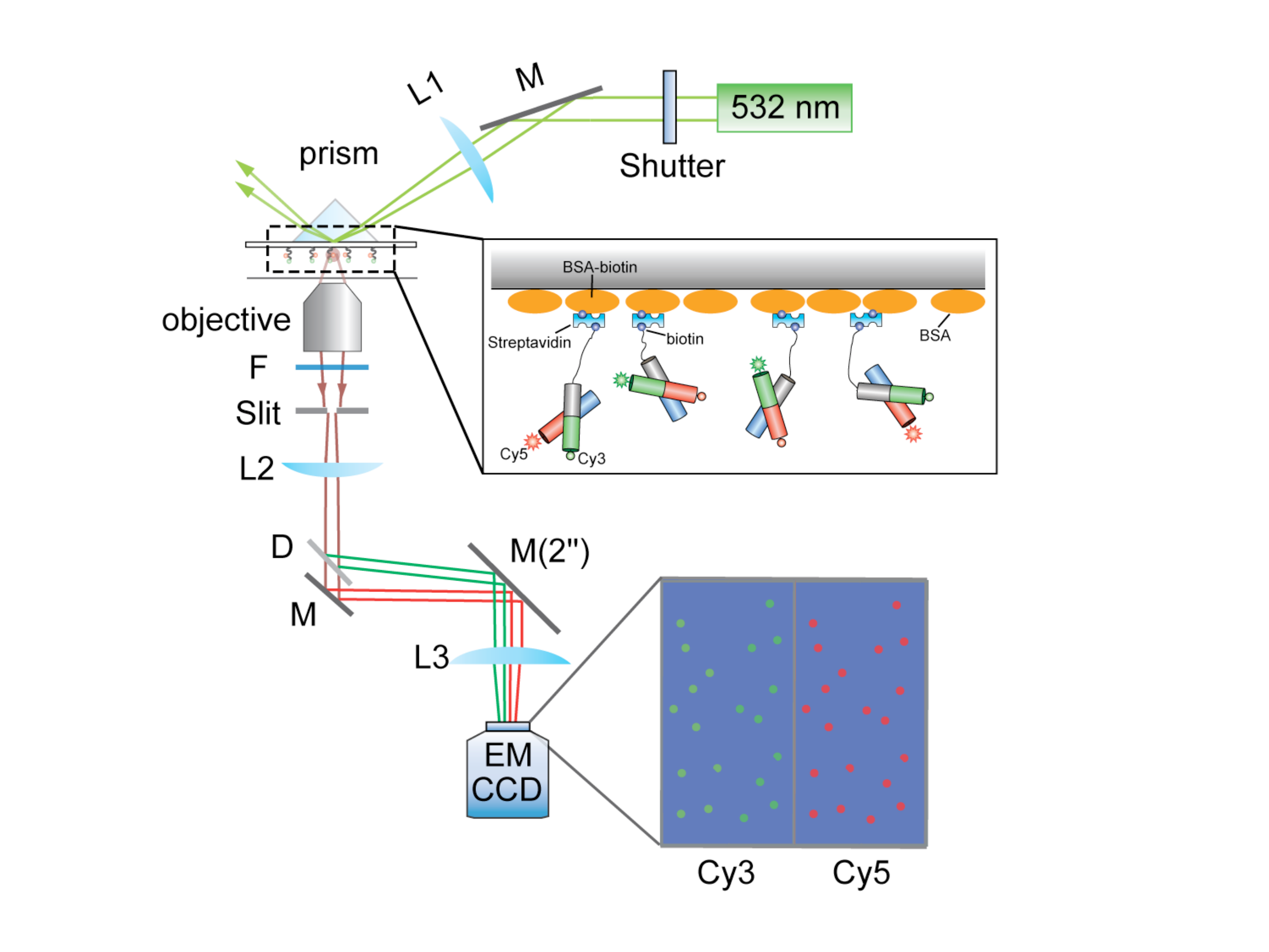}
\caption{A schematic diagram of experiments. DNA molecules were immobilized on a quartz surface via streptavidin-biotin interaction. Single-molecule images were taken in a prism-type TIRF (Total Internal Reflection Fluorescence) microscope. The fluorescence signals of Cy3 and Cy5 were collected by a water-immersion objective, and imaged on a separate areas of an EM-CCD camera (M : mirror, L : lens, F : filter, D : dichroic mirror).
\label{smFRET1_Fig}}
\end{figure}

\renewcommand{\thefigure}{S2}
\begin{figure}[tbp]
\includegraphics[width=6.0in]{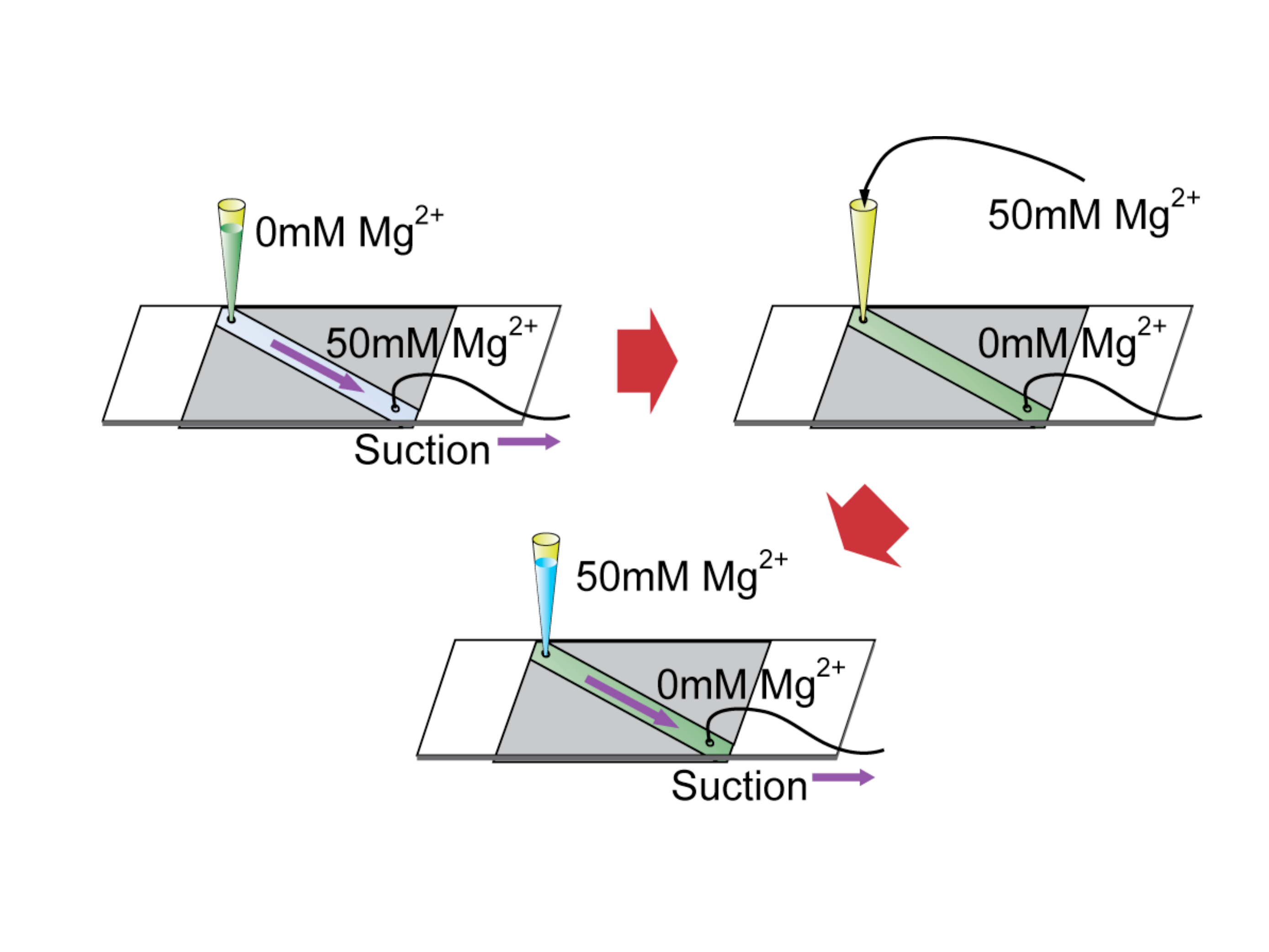}
\caption{ A schematic diagram of the magnesium pulse experiments. 
A sample chamber filled with 50 mM Mg$^{2+}$ was prepared, and a pipette tip with 0 mM Mg$^{2+}$ buffer inside was carefully plugged into a injection hole. 
While single-molecule images were taken, 0 mM Mg$^{2+}$ buffer was rapidly introduced into the detection chamber by pulling a syringe pump connected to the exit hole of the detection chamber via a Teflon tube. 
Remaining solution in the pipette tip was carefully replaced with a 50 mM Mg$^{2+}$ buffer, and the new solution containing 50 mM Mg$^{2+}$ was rapidly introduced into the detection chamber by suction. 
During the whole process, data acquisition was maintained.\label{smFRET_Fig2}}
\end{figure}

\renewcommand{\thefigure}{S3}
\begin{figure}[thp]
\includegraphics[width=6.5in]{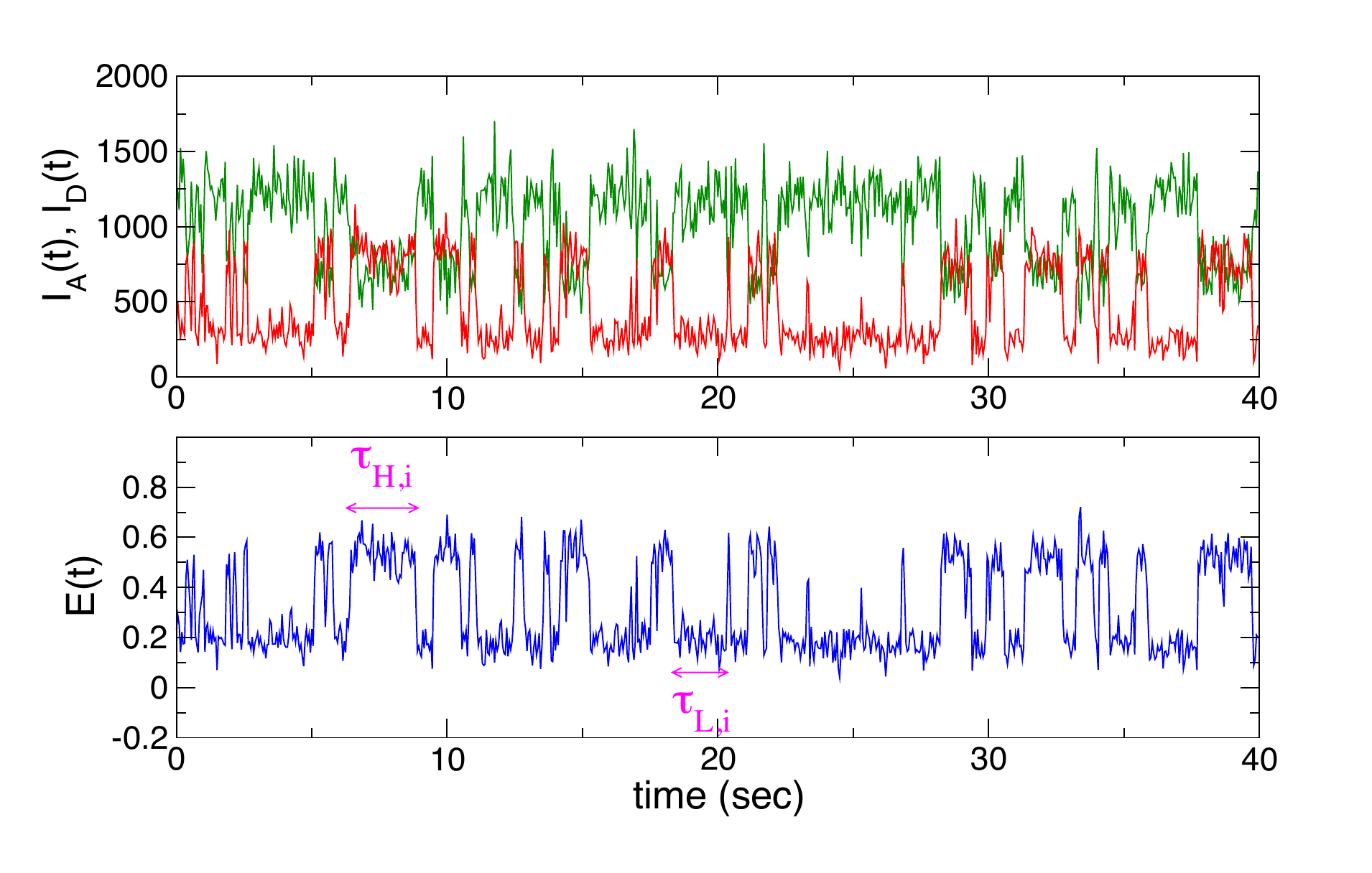}
\caption{(Top) Donor ($I_D$), acceptor ($I_A$) signals and (bottom) the corresponding FRET efficiency ($E(t)=I_A(t)/(I_A(t)+I_D(t))$) as a function of time.\label{Representative_trace}}
\end{figure}
\clearpage

\renewcommand{\thefigure}{S4}
\begin{figure}[hp]
\includegraphics[width=6.5in]{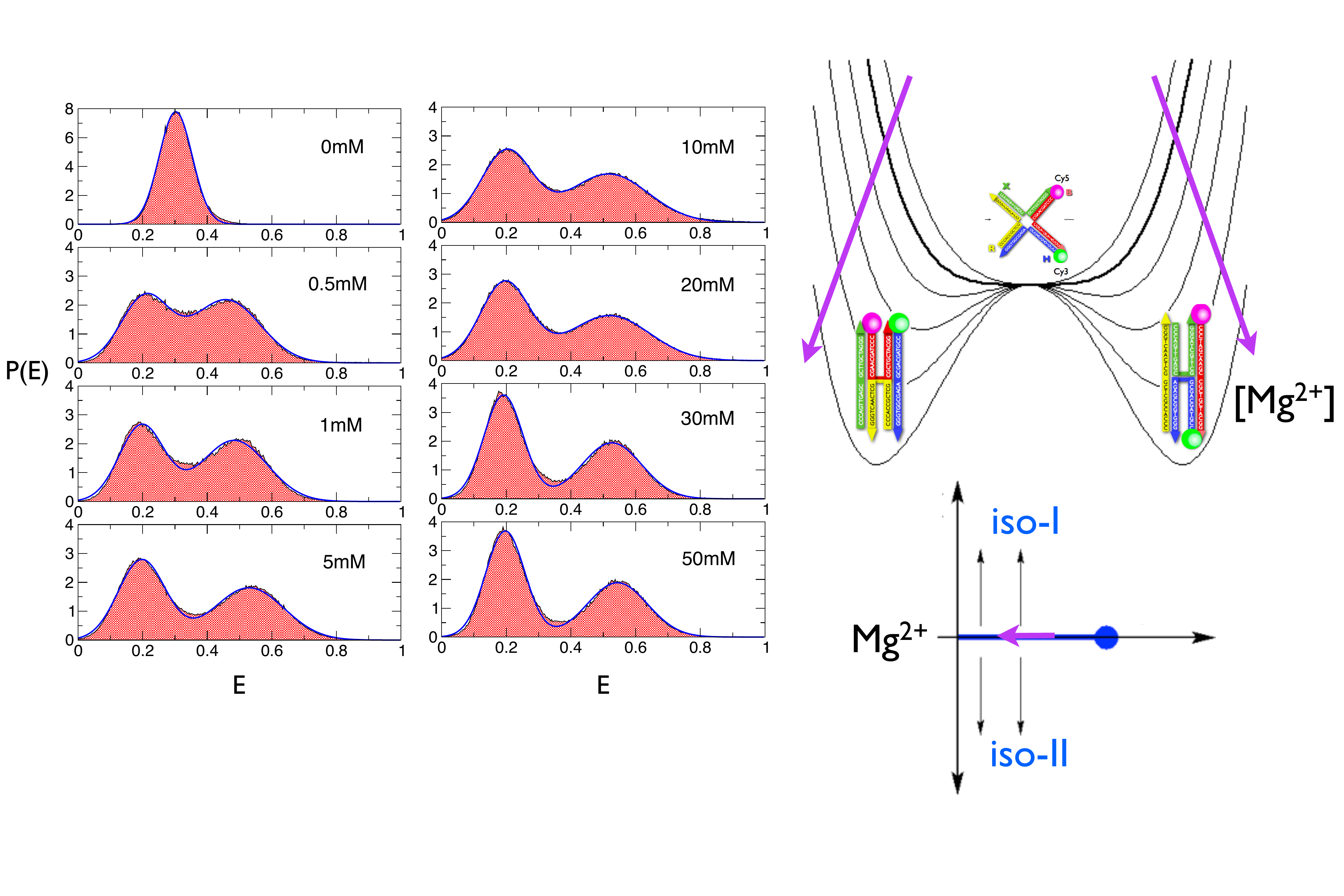}
\caption{Histogram of FRET value distribution with varying Mg$^{2+}$ concentrations. 
Each $P_{ens}(E)$ is fit to double-Gaussian distribution (blue line),  
$P_{ens}(E):=\phi \mathcal{G}_L(E)+(1-\phi)\mathcal{G}_H(E)$ where 
$\mathcal{G}_{\lambda}(E)=(2\pi\sigma^2_{E_{\lambda}})^{-1/2}e^{-(E-\overline{E}_{\lambda})^2/2\sigma^2_{E_{\lambda}}}$ ($\lambda=L,H$)
with the set of parameters shown in the box. 
Note that population of \emph{iso-I} and \emph{iso-II} are equally likely. Biomodal to uni-modal transition at a certain Mg$^{2+}$ ion concentration ($\le 50 \mu M$, see Ref.\cite{Joo04JMB}) is reminiscent of second order phase transition. 
In this case, Mg$^{2+}$ ions is analogous to inverse temperature in an Ising spin system (see Mg$^{2+}$ ion dependent free energy and phase diagram on the right). 
\label{Distribution}}
\end{figure}
\centering
\begin{tabular}{|r||r|r|r|r|r|}
\hline
[Mg$^{2+}$] &$\phi$ &  $\overline{E}_L$ & $\sigma_{E_L}$ & $\overline{E}_H$& $\sigma_{E_H}$ \\
\hline
0mM    &  1      & 0.30 & 0.051 & - & -\\
0.5 mM & 0.42 &0.21 &  0.074 &  0.46 & 0.106 \\
1mM    & 0.46 & 0.20 & 0.069 &  0.48 &  0.102 \\
5mM    & 0.51 & 0.20 & 0.073 &  0.53 &  0.108  \\
10mM   & 0.48 & 0.20 & 0.078 & 0.52 & 0.123\\
20mM   & 0.51 & 0.20 & 0.075 & 0.52 &  0.125  \\
30mM   & 0.45 & 0.19 & 0.061 & 0.53 & 0.093  \\ 
50mM   & 0.55 & 0.20 & 0.061 & 0.54 & 0.093  \\
\hline
\end{tabular}

\renewcommand{\thefigure}{S5}
\begin{figure}[hp]
\includegraphics[width=6.0in]{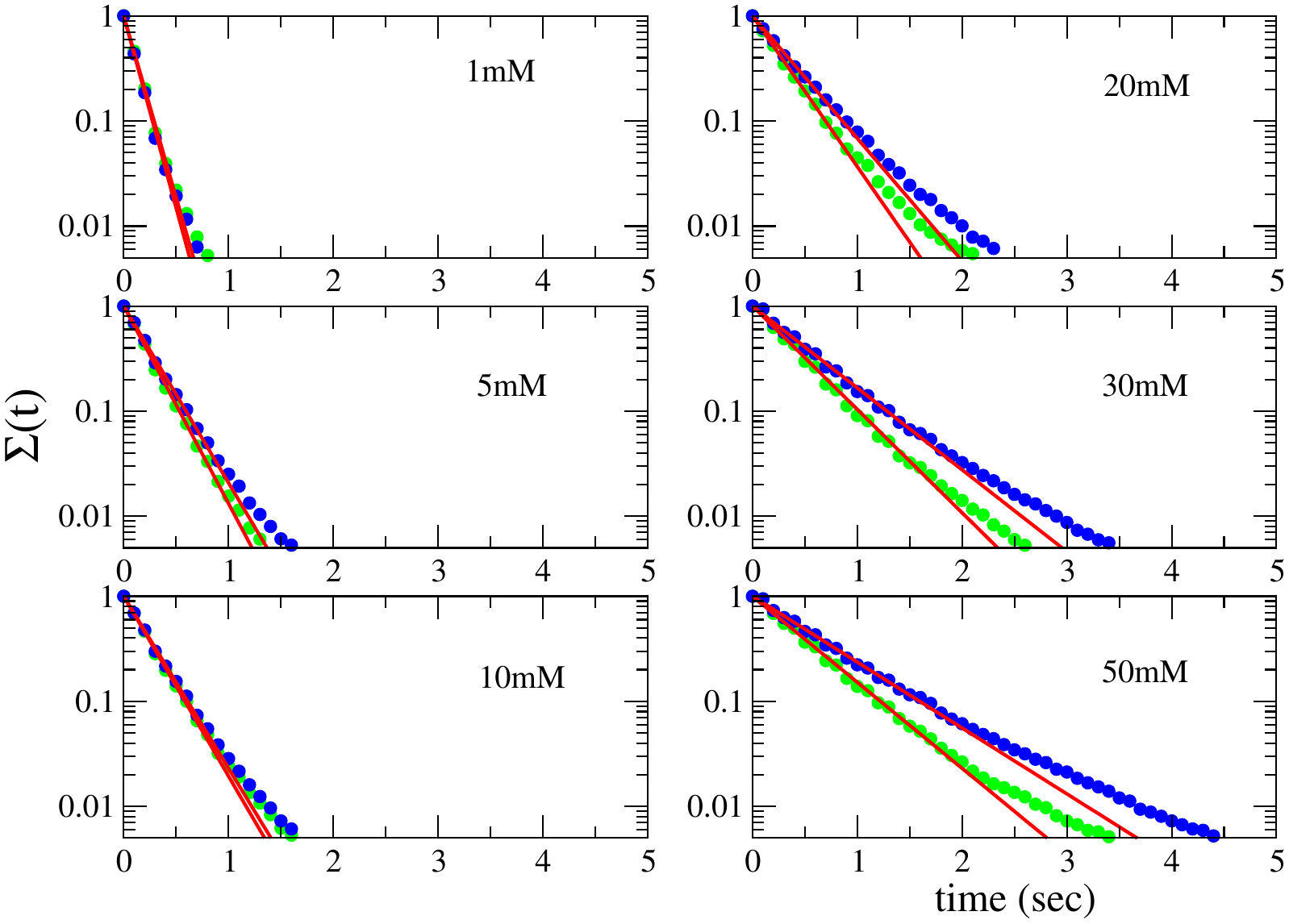}
\caption{The probability of remaining in high (green) and low $E$ values (blue). Red lines correspond to single exponentials.  
The mean life time in each state can be obtained using $\langle\tau\rangle=\int^{\infty}_0d\tau \tau p_{dwell}(\tau)=\int^{\infty}_0d\tau \Sigma (\tau)$ where $\Sigma(t)=1-\int^t_0d\tau p_{dwell}(\tau)$. \label{Dwell_analysis}}
\end{figure}

\renewcommand{\thefigure}{S6}
\begin{figure}[tbp]
\centering
\includegraphics[width=4.0in]{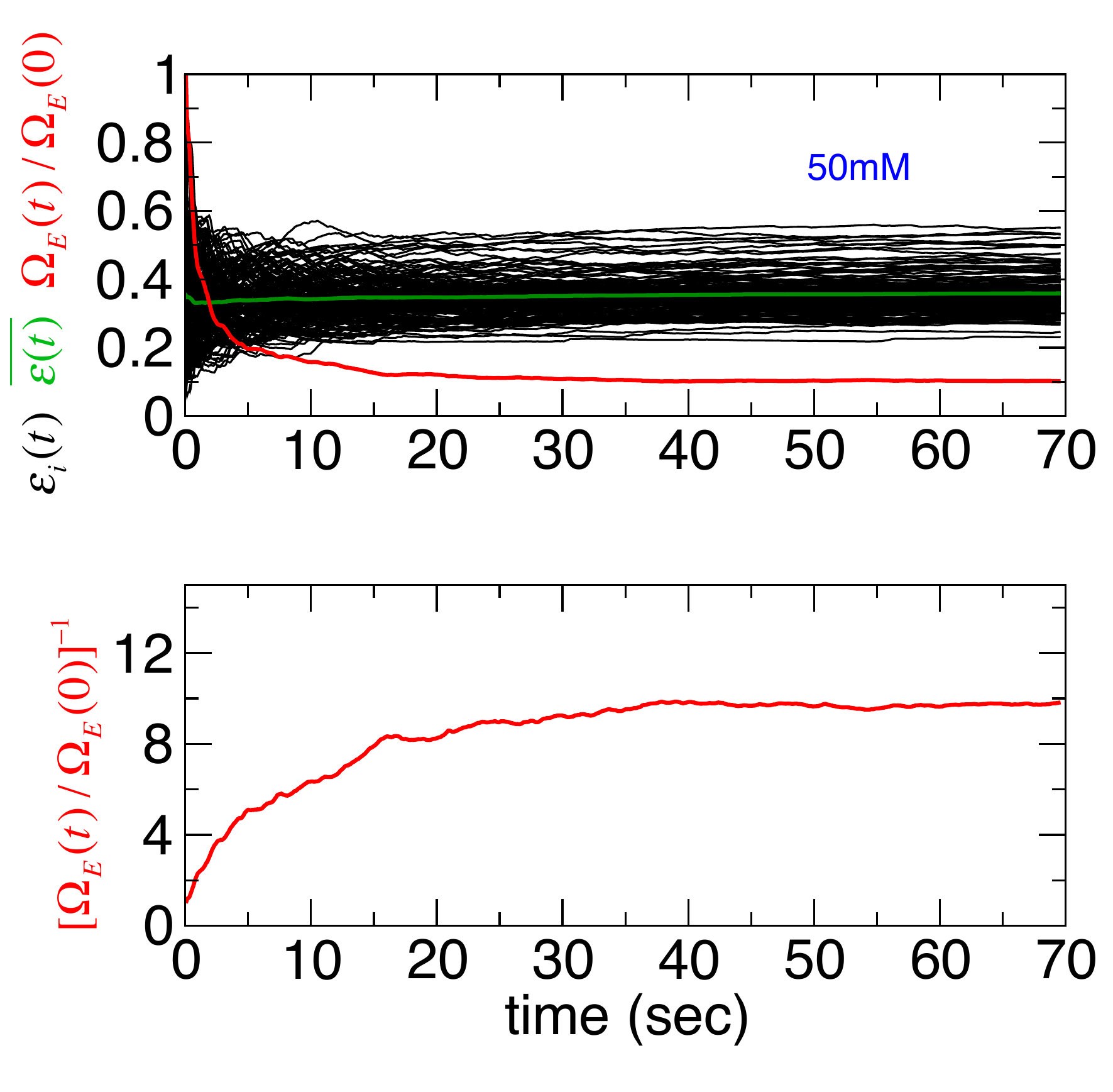}
\caption{Analysis of the transition dynamics of HJ trajectories at [Mg$^{2+}$]=50 mM shows that the ergodicity is still broken over an extended observation time $\mathcal{T}_{obs}\approx 70$ sec. $\varepsilon_{\i}(t)$, $\overline{\varepsilon(t)}$ and $\Omega_E(t)/\Omega_E(0)$ (or $[\Omega_E(t)/\Omega_E(0)]^{-1}$) are shown in grey, green, and red lines, respectively. 
 \label{graph_Tave_50mM_long}}
\end{figure}

\renewcommand{\thefigure}{S7}
\begin{figure}[tbp]
\includegraphics[width=6.0in]{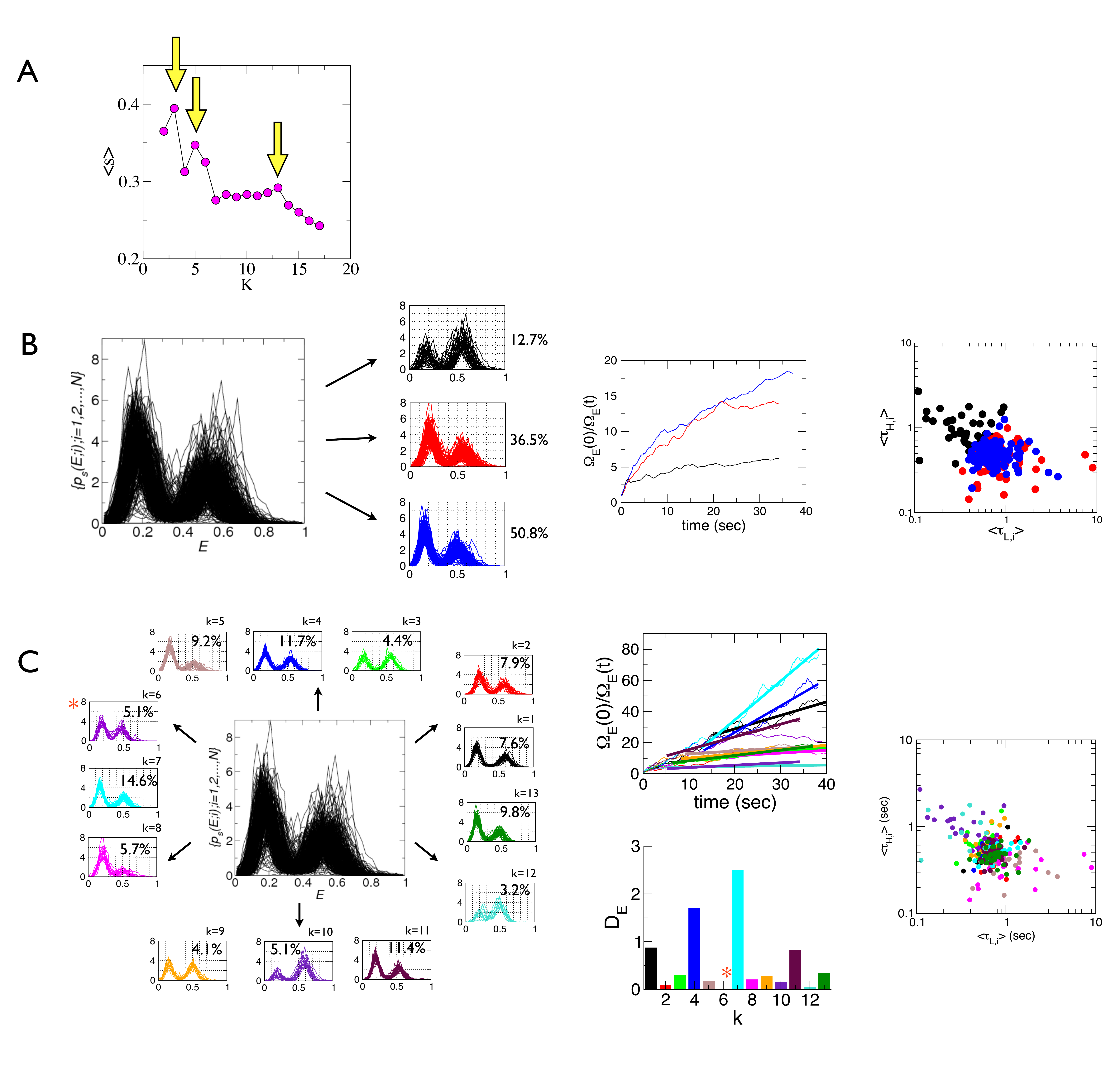}
\caption{Clustering analysis of FRET trajectories at [Mg$^{2+}$]=50 mM. {\bf A.} Mean silhouette function ($\langle s\rangle$) as a function of the number of clusters, K. 
Optimal $\langle s\rangle$ values are found at K=3, 5, 13 as indicated by the arrows.   
{\bf B.} Partitioning of $\{p_s(E;i)\}$ with K=3 yields three clusters (k=1 (12.7 \%), k=2 (36.5 \%), k=3 (50.8 \%)) but the dynamics within the cluster, k=2, is still non-ergodic. 
$\Omega_E(0)/\Omega_E(t)$ and scatter plot of ($\langle \tau_{L,i}\rangle$,$\langle\tau_{H,i}\rangle$), partitioned into three clusters, are shown on the right.  
{\bf C.} Partitioning of $\{p_s(E;i)\}$ with K=13 are shown. The cluster k=6 is still non-ergodic but its percentage (5.1 \%) is small.   
  $\Omega_E(0)/\Omega_E(t)$, $D_E$ calculated for each $k(=1,2\ldots,13)$, and scatter plot of ($\langle \tau_{L,i}\rangle$,$\langle\tau_{H,i}\rangle$), partitioned into 13 clusters, are shown on the right. 
\label{silhouette}}
\end{figure}

\renewcommand{\thefigure}{S8}
\begin{figure}[tbp]
\includegraphics[width=6.0in]{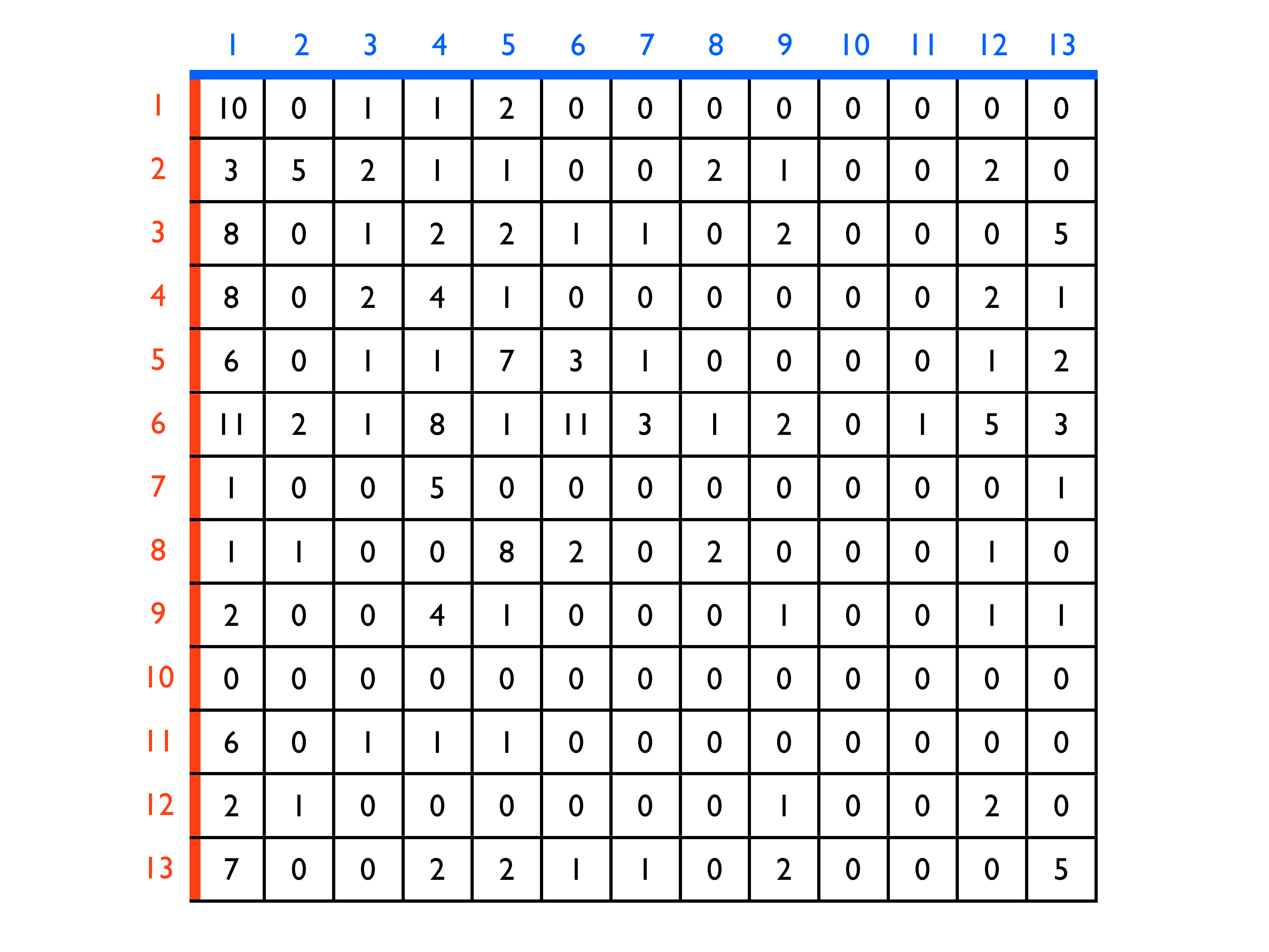}
\caption{Mg$^{2+}$ pulse-induced transition frequency matrix among 13 kinetically disjoint states based on 148 FRET trajectories. 
The indexes at the sides of matrix denote the cluster number $k=1,2\ldots 13$. 
The number in the matrix element represents the number of transitions from state $\xi$ to state $\eta$ induced by the sequence of Mg$^{2+}$ pulses (see Fig.S2 and Fig. 6a). 
\label{transition_matrix_13}}
\end{figure}

\renewcommand{\thefigure}{S9}
\begin{figure}[tbp]
\includegraphics[width=6.0in]{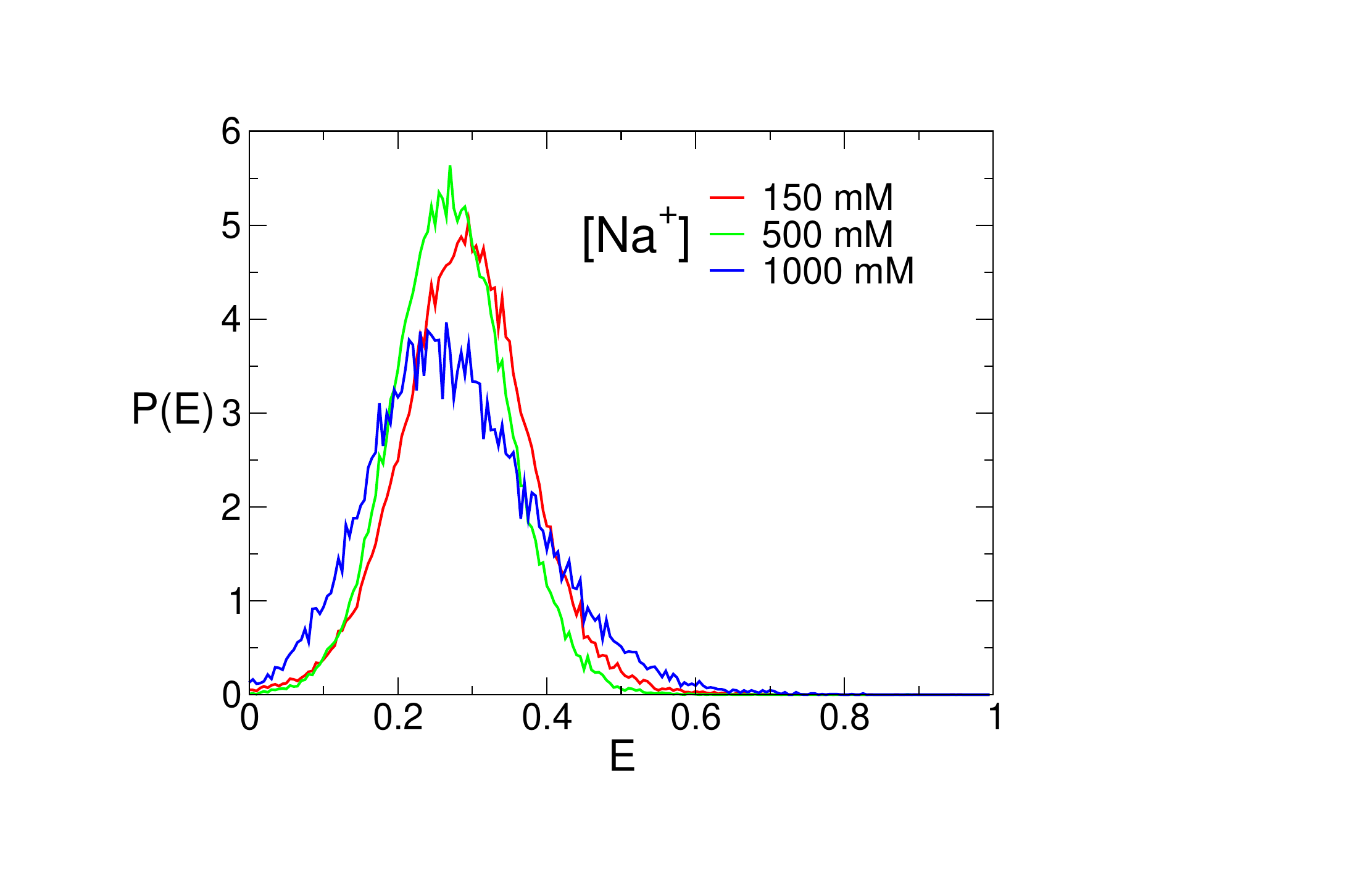}
\caption{All the histograms of FRET efficiency, $P(E)$, under 150, 500, 1000 mM monovalent salt condition using NaCl but in the absence of Mg$^{2+}$ ions are uni-modal, suggesting that monovalent ions do not induce the two-state like isomerization dynamics in HJs. 
This control experiment suggests that isomerization dynamics and heterogeneity of this dynamics are related to specific DNA-Mg binding events, not the screening effects due to the high ionic strength in 50 mM MgCl$_2$ condition.         
\label{NaCl}}
\end{figure}

\renewcommand{\thefigure}{S10}
\begin{figure}[tp]
\includegraphics[width=6.0in]{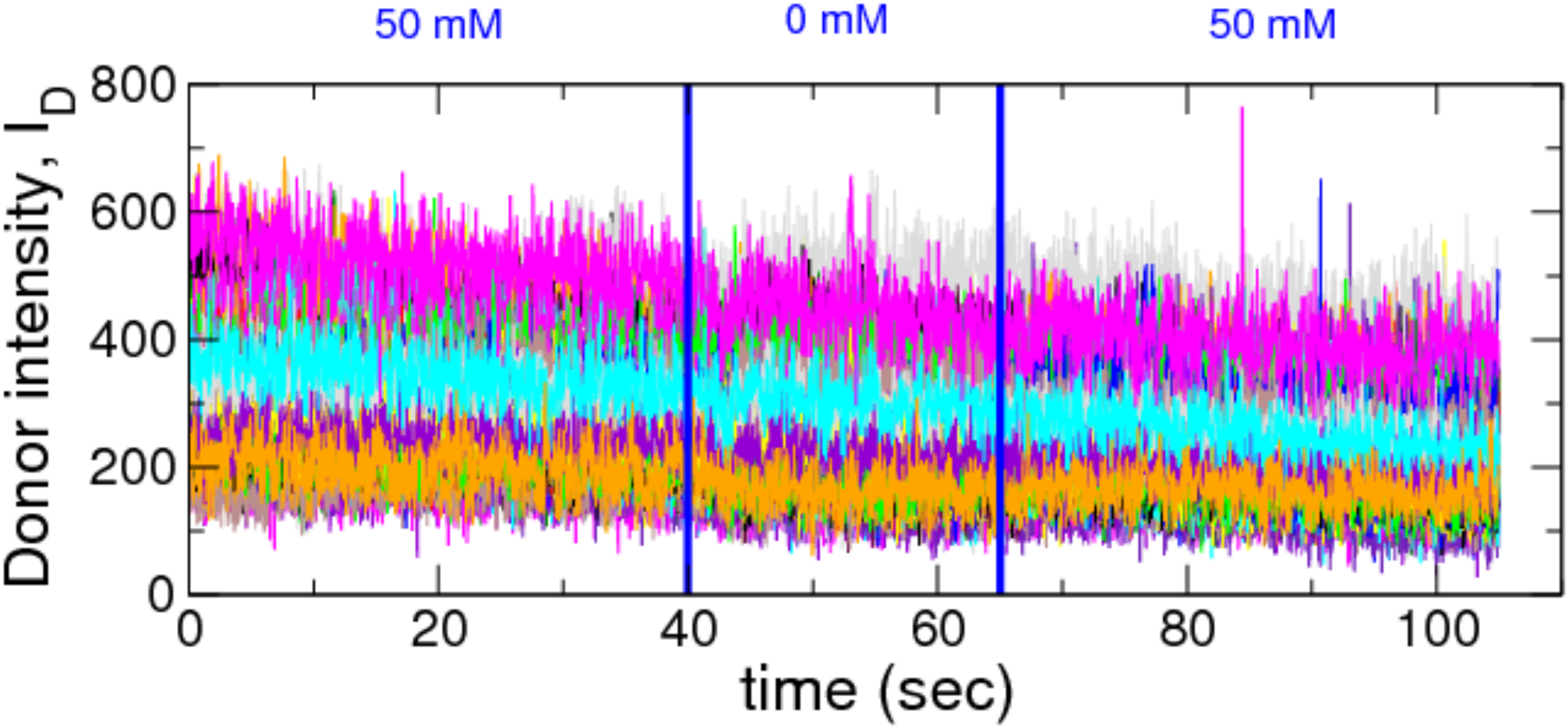}
\caption{Control experiment $-$ Mg$^{2+}$-pulse experiments using donor-only-tagged HJs. The Mg$^{2+}$ is switched at 40 s (50 mM $\rightarrow$ 0 mM) and 65 sec (0 mM $\rightarrow$ 50 mM). It is of note that although the photon intensity itself varies from molecule-to-molecule and there are weak gradual drift in the data, the overall intensities from the donor dyes are constantly maintained irrespective of the Mg$^{2+}$ concentration. 
It is hard to see the change in the pattern of data as in Fig.6a when only the donor dyes are tagged to the molecule. 
This observation reinforces our hypothesis that the HJ dynamics observed with sm-FRET are originated from the conformational dynamics of HJs, not due to the dye-nucleotide stacking or heterogeneous interaction with surface.    
\label{control}}
\end{figure}

\end{document}